\documentclass[acmlarge,screen]{acmart}  % ,review

\usepackage{setspace}
\usepackage{amsmath}
\usepackage{hyperref}

\usepackage{subfigure}

\usepackage{algpseudocode} 
\usepackage{multirow}
\usepackage{enumerate}
\usepackage[figuresright]{rotating}
\newcommand{\red}[1]{\textcolor{black}{#1}} %revision note
\usepackage{mdframed}
\usepackage[ruled,linesnumbered]{algorithm2e}
\usepackage{tablefootnote}
\AtBeginDocument{%
  \providecommand\BibTeX{{%
    \normalfont B\kern-0.5em{\scshape i\kern-0.25em b}\kern-0.8em\TeX}}}

\setcopyright{acmcopyright}
\copyrightyear{2022}
\acmYear{2022}
\acmDOI{XX.XXXX/xxxxxxx.xxxxxxx}

\acmConference[SIGMETRICS '23]{SIGMETRICS '23: ACM SIGMETRICS conference}{June 19--23, 2023}{Orlando, Florida, USA}
\acmBooktitle{Proceedings of the ACM SIGMETRICS Conference 2023
(SIGMETRICS’23), June 19--23, 2023, Orlando, Florida, USA}
\acmPrice{15.00}
\begin{document}

\title[Money Laundering]{Towards Understanding Crypto Money Laundering in Web3 Through the Lenses of Ethereum Heists} 
%Towards Understanding Cryptocurrency Money Laundering: Cases Studies of Ethereum Heists


%\author{Anonymous author(s)}

\author{Dan Lin, Jiajing Wu*, Qishuang Fu, Yunmei Yu, Kaixin Lin, Zibin Zheng, Shuo Yang}

% \author{Dan Lin}
% %\authornote{Both authors contributed equally to this research.}
% \orcid{0000-0001-7067-2396}
% %\author{G.K.M. Tobin}
% %\email{webmaster@marysville-ohio.com}
% \affiliation{%
%   \institution{School of Software Engineering, \\Sun Yat-sen University}
%   \city{Zhuhai}
% \email{lind8@mail2.sysu.edu.cn}

% \author{Jiajing Wu}
% \authornote{Corresponding author.}
% \affiliation{%
%   \institution{School of Computer Science and \\Engineering, Sun Yat-sen University}
%   \city{Guangzhou}
% \email{wujiajing@mail.sysu.edu.cn}

% \author{Qishuang Fu}
% \affiliation{%
%   \institution{School of Computer Science and \\Engineering, Sun Yat-sen University}
%   \city{Guangzhou}
% \email{fuqsh6@mail2.sysu.edu.cn}

% \author{Tao Huang}
% \affiliation{%
%   \institution{School of Computer Science and \\Engineering, Sun Yat-sen University}
%   \city{Guangzhou}
% \email{huangt97@mail2.sysu.edu.cn}

% \author{Zibin Zheng}
% \affiliation{%
%   \institution{School of Software Engineering, \\Sun Yat-sen University}
%   \city{Zhuhai}
% \email{zhzibin@mail.sysu.edu.cn}

\renewcommand{\shortauthors}{Anonymous author(s)}

% % 删除多余的reference信息
% % https://www.twblogs.net/a/5f020cc7d496dddbb5425051/?lang=zh-cn
% \settopmatter{printacmref=false} 
% \renewcommand\footnotetextcopyrightpermission[1]{}
\begin{abstract}
% zj comment 第一行右边超出边界了【get】
With the overall momentum of the blockchain industry, crypto-based crimes are becoming more and more prevalent. After committing a crime, the main goal of cybercriminals is to obfuscate the source of the illicit funds in order to convert them into cash and get away with it. Many studies have analyzed money laundering in the field of the traditional financial sector and blockchain-based Bitcoin. But so far, little is known about the characteristics of crypto money laundering in the blockchain-based Web3 ecosystem. To fill this gap, and considering that Ethereum is the largest platform on Web3, in this paper, we systematically study the behavioral characteristics and economic impact of money laundering accounts through the lenses of Ethereum heists. Based on a very small number of tagged accounts of exchange hackers, DeFi exploiters, and scammers, we mine untagged money laundering groups through heuristic transaction tracking methods, to carve out a full picture of security incidents. By analyzing account characteristics and transaction networks, we obtain many interesting findings about crypto money laundering in Web3, observing the escalating money laundering methods such as creating counterfeit tokens and masquerading as speculators. Finally, based on these findings we provide inspiration for anti-money laundering to promote the healthy development of the Web3 ecosystem.

\end{abstract}



% https://blog.csdn.net/weixin_41971366/article/details/107722049

% \keywords{Money laundering, Cryptocurrency, Cybercriminal, Blockchain, Network analysis}

\maketitle

% 注意顺序

\section{Introduction}

% 【P1区块链以及基于区块链的虚拟货币和web3生态，在过去十年间获得了快速的发展。
% 具体可以展开一下比特币，整个虚拟货币市场的市值达到一个惊人的规模。然后web3包括什么。加入原P2的部分内容。但是这一段不要太长。】

% 在过去十年间见证了区块链以及基于区块链的加密货币生态的快速发展。加密货币的市值达到一个惊人的规模，其中比特币市值高达$385 Billion (CoinMarketCap)。同时，随着区块链技术的进一步发展，全球掀起了互联网的第三次迭代（Web3）的浪潮。Web3的颠覆性建立在三个基本要素：存储交易记录并确保Web3的去中心化性质的底层区块链、代表应用逻辑的智能合约、可以代表任何有价值的东西的数字资产。Web3上共享、共建、可组合的经济系统带来了比传统金融和公链生态更丰富的应用生态、更开放的经济体系以及更庞大的交易量。
% zj comment: three essential fundamentals这句话，要不要用分点符号 blabla :1)...;2)...;3)...
The past decade has witnessed the rapid growth of blockchain and the blockchain-based cryptocurrency ecosystem. The market capitalization of cryptocurrencies has reached a staggering scale, with Bitcoin reaching a market capitalization of \$385 Billion~\cite{Chainalysis2022Web3}. Meanwhile, with the further development of blockchain technology, there is a global wave of the third iteration of the Internet (Web3).
Web3's disruption is built on three essential fundamentals~\cite{McKinsey2022Web3}: an underlying blockchain that stores transaction records and ensures the decentralized nature of Web3, smart contracts that represent the logic of the application, and crypto assets (also called digital assets) that can represent anything of value. The shared, co-constructed, assemblable economic system on Web3 brings a richer application ecosystem, a more open economic system, and a larger transaction volume than traditional financial and public blockchains.

However, any new technology, especially those with a lack of regulation, can be exploited for unscrupulous gain. Since blockchain transactions do not require user identification information, blockchain and its ecosystem have become a hotbed of various cybercrimes and illegal financial activities~\cite{Wu2021JNCA}, and the still-developing blockchain-based Web3 is no exception.
%However, any new technology can be misused by the bad guys for personal gain, and the still-developing Web3 is no exception. 
% ys comment 第一段没有点名区块链和Web3的关系，第二段就直接说区块链的问题和Web3存在，这点有些奇怪，可以说明一下Web3的底层支持是区块链，区块链保障了Web3去中心化的性质，才使得区块链下存在的一些问题在Web3下也亟待解决 【有道理，在第一段点明了Web3的基本要素之一“an underlying blockchain that stores transaction records and ensures the decentralized nature of Web3”】
% mingxi's comment: “bad guys"听起来有点怪，是不是有点口语化；这里说明still-developing好像有点突兀而且对paper好像没有什么用【bad guys换成了unscrupulous people】
According to blockchain security firm Certik~\cite{certik2022Web3}, in the first half of 2022 alone, more than \$2 billion was stolen from Web3 projects as a result of hacking and vulnerabilities. After stealing crypto assets, cybercriminals conceal and disguise them through different channels to make them appear legitimate and then withdraw them, a process known as money laundering. So it is said that money laundering is the subsequent part of all other forms of crypto-based crimes~\cite{Chainalysis2021,Chainalysis2022}. 
% ys comment 感觉After到 a process known这一句可以rephrase一下，读起来有点怪怪的，没有主语【已改正】
% mingxi's comment: "so it is said that xxx", who said that? more citations are needed. 【已添加引用】
Therefore, with the frequent occurrence of Web3 security incidents, crypto anti-money laundering (AML) is in a crucial position to be the last line of defense to stop hackers from successfully cashing out and also to deter hackers from committing Web3 crimes at the same time.
%Therefore, with the frequent occurrence of Web3 security incidents, anti-money laundering is in a crucial position as the last line of defense to stop hackers from successfully cashing out, and can in turn play a role in deterring hackers from committing crimes.

Anti-money laundering (AML) is not a new issue, and a wealth of research on the AML issue in traditional financial scenarios have been proposed~\cite{Gao2007data,Wronka2021Cyber,Dumitrescu2022Bank,Rocha-Salazar2021,fatf2022}. In the field of cryptocurrency, the Elliptic dataset~\cite{elliptic2019Weber} is the first open-source Bitcoin money laundering dataset that labels abnormal/normal Bitcoin transactions roughly. But for one, this dataset only has binary labels for money laundering with no other business details, and for two, the Bitcoin platform this dataset focuses on is very different from the Web3 ecosystem which contains rich decentralized applications (DApps).
%, but not enough to represent the Web3 ecosystem containing rich decentralized applications (DApps). 
% ys comment 上面说not enough其实就是说Web3的大体量和丰富性，传统无法解决，这个特点应该在第一段第二段点明，让读者明白Web3有什么特点 【有道理，前面已补充】
To the best of our knowledge, there is currently no public dataset on Web3 money laundering in academia, nor is there a systematic description and analysis of money laundering on the Web3 ecosystem. It is not clear what the transaction characteristics of these Web3 money laundering accounts are, how the flow of illegal funds in the money laundering ring has achieved the effect of obfuscating the source, and what kind of impact it has on the economy of the Web3 ecosystem. Therefore, this is the question that this paper wants to explore.

%Trying to understand crypto money laundering in Web3 is challenging, and the challenges arise from three main levels:
However, due to the unique characteristics of Web3 and money laundering practices on it, AML approaches on traditional financial scenarios or bitcoin cannot be directly applied to Web3 due to the following three challenges.
%The challenge of understanding crypto money laundering in Web3 comes from three main levels.
    \textbf{(i)} The underlying blockchain. Compared with traditional finance, blockchain is decentralized, borderless, and anonymous, without limiting the number of accounts each user can create. This allows cybercriminals to conduct a large number of frequent transactions between accounts under their control, leading to difficult identification of account entities and a large number of anonymous transfers. 
%     This challenge also exists in Bitcoin AML research. The following two challenges are specific to Web3.
    \textbf{(ii)} Smart contracts and digital assets. Based on blockchain, smart contracts enable various types of digital assets which can be exchanged in the trading platforms. At the same time, Turing-complete smart contracts can represent and execute more complex application logic and functions, leading to more complex transaction patterns. % (ii) 智能合约及数字资产。通过智能合约可实现各类的数字资产，这些不同数字资产之间可以进行兑换。同时，图灵完备的智能合约可以定义和执行更加复杂的业务逻辑和功能，导致了更为复杂的交易模式。
    % ys comment smart contract这一段逻辑有点怪，正确的逻辑应该是：合约承载assets，实现了复杂的交易逻辑=>所以合约变得更复杂=>所以有更加复杂的交易pattern，问题很复杂【已修改，由于合约可以实现复杂的逻辑导致了更复杂的交易模式】
    \textbf{(iii)} Decentralized finance (DeFi)~\cite{werner2021sok}. On the one hand, immature DeFi applications gather a large number of assets, attracting the attention of criminals and becoming the hardest hit by asset theft; on the other hand, DeFi services lacking anti-money laundering compliance bring ever-changing means of exchanging coins while also fueling crypto criminals to launder dirty money.

% \noindent $\bullet$
    % 包括稳定币，同质化代币，NFT等等，, including stablecoins~\cite{Moin2020Stablecoin}, ERC20 tokens~\cite{Chen2020ERC20}, Non-Fungible Token (NFTs)~\cite{Bamakan2022NFT}, etc
    
In this paper, we go for the first time to characterize and analyze the crypto money laundering behavior in Web3, taking the largest blockchain platform of Web3~\cite{Chainalysis2022Web3}, Ethereum, as an entry point. Note that only the information of the accounts where the security incidents occurred is publicly reported, whereas the money laundering accounts where the stolen money is transferred are usually unknown. To this end, we start from the tip of the iceberg - a very small number of accounts of known security incidents - and then dig and expand the malicious addresses of money laundering, in order to carve out the full picture of the security incidents and complete the chain of evidence for the transfer of stolen assets. 
% zj comment 这个句子也有点长，考虑把"in order to carve out"这句单列出来？
% 请注意，目前被开源的只有安全事件发生的账户信息，而后进行赃款转移的洗钱账户信息并不知晓。本文要做的事情就是从这些已知的冰山一角的安全事件源头出发，挖掘并扩充后续恶意地址标签，对安全事件的全貌进行刻画，完善案件被盗资产转移证据链。

Specifically, we first propose an abstract model to describe the process of money laundering and present a heuristic tracing algorithm based on this abstract model to extract money laundering transactions from the massive amount of anonymous blockchain data (\textbf{Section~\ref{sec:study_design}}). We construct the \emph{first money laundering dataset} (containing over 160,000 addresses) in Web3, called $\textsf{EthereumHeist}$, and also use a case study to illustrate the effectiveness of the tracing method. % the complexity of money laundering research and 
With these real data, we conduct in-depth empirical analysis from micro to macro perspectives.
(i) From the perspective of individual laundering accounts, we count and analyze what are the characteristics of accounts and their transaction behavior in the money laundering process (\textbf{Section~\ref{sec:accounts}}). 
(ii) From the perspective of a gang, we model and measure the network of transactions involved in the cases and analyze the difference between money laundering transaction networks and the entire Ethereum transaction network (\textbf{Section~\ref{sec:network_properties}}). 
(iii) From a more macro perspective, we explore how the flow of money laundered funds affects the economy of the Web3 ecosystem (\textbf{Section~\ref{sec:economic}}). Finally, we discuss the AML implications, limitations, and ethical issues of this paper, as well as possible directions for future research. \textbf{The main contributions are as follows.}
% $\bullet$ 
\begin{itemize}
    \item To our best knowledge, we present the \emph{first} dataset and the \emph{first} systematic analysis on crypto money laundering in Web3 through the lenses of Ethereum heists from 2018-2022. Based on a very small number of tagged accounts of hackers, exploiters, and scammers, we adopt an augmented poison policy to trace the untagged money laundering process, which provides a full picture of incidents. Methods for tracing, data collection, and measurement of $\textsf{EthereumHeist}$ can also be reused for other cases. We present the money laundering dataset which can be found at \url{https://www.dropbox.com/}. %, so as to build a more complete crypto transaction tracing system in Web3  heuristic tracing
    % zj comment 感觉最后一句话之前可以加一个"我们提出了洗钱数据集，“然后在说the data can be found in httpxxx。因为数据集也是一个比较好的贡献。【get】
     
    \item We obtain many interesting findings of crypto money laundering in Web3 by adopting feature analysis, graph analysis, and other methods. These findings help us gain new knowledge about the crypto money laundering behaviors in Web3. Particularly, we find that it is common for exploiters to obfuscate stolen funds by swapping tokens through DeFi platforms, and hackers even launder money by creating counterfeit tokens for higher anonymity.  

    \item We conduct an empirical study to understand the economic impact of laundering in the Web3 ecosystem by investigating the evolution of money laundering destination service providers and the market price of crypto assets. Moreover, we present insights for anti-money laundering in the Web3 ecosystem based on trends of money laundering techniques and service providers, in order to promote the healthy development of Web3.
    
\end{itemize}

\section{Background}

%\vspace{-1ex}

% \subsection{The Building Blocks of Web3}

\subsection{Stolen Funds From Web3}

% 加密货币生态系统的繁荣推动了对交易平台的需求。但是哪里有钱，哪里就吸引盗贼的目光。
% 中心化交易所聚集大量资金且多数情况下防御脆弱，一直是黑客觊觎的对象。从Mt.Gox被盗至今发生54起交易所被盗事件（Cryptosec）.
% 还处在发展初期的DeFi近年来也是黑客攻击的首选目标，DeFi数字资产被盗主要是因为合约漏洞、闪电贷攻击、私钥泄露等原因。据报道，2020年以来有75起DeFi漏洞发生，损失约为17亿美元（Cryptosec）。
% 更普遍但缺乏充分披露的资产被盗类型是欺诈。黑客通过恶意邮件或虚假宣传对加密货币个人持有资产实施盗窃，如钓鱼诈骗、旁氏骗局、拉地毯骗局等。
% Etherscan上标记了部分盗窃以太坊加密资产的账户，标签为“heist”。

The boom in the Web3 ecosystem is driving demand for trading platforms. However, where there is money, that is where thieves are attracted. The sources of stolen funds on Web3 can be broadly classified into three types.
\textbf{Centralized exchanges} (CEXes), which gather large amounts of money but in most cases have weak defenses, have been coveted by hackers. %Since the MtGox exchange theft, there have been 54 exchange thefts.
\textbf{DeFi projects}, which are still in the early stage of development, have also been a prime target for hackers in recent years, with DeFi digital assets stolen mainly due to contract vulnerabilities, flash loan attacks, and private key leaks. %There have been 104 reported DeFi breaches since 2020, with losses of approximately \$3.6 billion\footnote{\url{https://cryptosec.info/defi-hacks/}}.
\textbf{Scams} are a more common but under-disclosed type of asset theft. Scammers commit theft of cryptocurrency personal holdings through malicious emails or false propaganda, such as phishing scams, Ponzi scams, etc.

% mingxi's comment: 第三段的\textbf{}应该还是可以想办法放到句首的，这样整齐一点【有道理，已修改】
% zj comment: 感觉CEX，DeFi 和Scams不是一个并列关系。CEX和DeFi这两个词只蕴含了中立的对象，Scams蕴含了手段+对象（诈骗手段+个体户受害者）的意思。如果要并列的话应该是把scams换成individuals。不知道有没有说清楚:(

%\vspace{-1.5ex}

\subsection{Money Laundering}

\label{subsec:moneylaundering}
% 洗钱是转移不义之财或犯罪活动（如贩毒或资助恐怖分子等）产生的资金，以掩盖和隐藏资金来源的非法过程。
% 洗钱通常包括三个主要阶段:
% (i)放置：将犯罪收益投入“清洗系统”的过程。这些非法收益常被分成较小的金额放在多个账户中，以防止被反洗钱系统发现。 
% (ii)分层：通过复杂的多种、多层的金融交易，将非法收益与其来源分开，并进行最大限度的分散，以掩饰线索和隐藏身份。转移的频次越高，调查人员想要通过网络路径追查源头的难度就越大。
% (iii) 整合（Integration）：是洗钱的最后阶段，在整合阶段，被形象地描述为“甩干”。资金被整合到金融系统中，就像它们是合法的一样。
Money laundering is the illegal process of transferring funds generated by ill-gotten gains or criminal activities (such as drug trafficking or terrorist financing) in order to conceal and hide the source of the funds. %The money from the criminal activity is considered dirty, and the process “launders” it to make it look clean. 
Money laundering typically consists of three main stages: 
(i) \textbf{Placement}: The process of putting the proceeds of crime into the ``laundering system''. These illicit proceeds are often divided into smaller amounts and placed in multiple accounts to prevent detection by AML systems. 
(ii) \textbf{Layering}: Separation of illicit proceeds from their sources and maximum dispersion through complex multiple, multi-layered financial transactions to disguise leads and hide identities. The higher the frequency of diversion, the more difficult it is for investigators to trace the source through network paths.
%draining 这个词不是很懂
(iii) \textbf{Integration}: This is the final stage of money laundering, which is graphically described as ``draining''. The funds are integrated into the financial system as if they were legitimate.
% --------------
% mingxi‘s comment：冒号后面的单词貌似一般小写 & 冒号后的各个元素之间貌似用分号；不是特别确定\textbf{} + 句子的情况，建议查一下确认没有writing的问题【感谢铭熙指出~ 我查了一下，英语冒号后面如果是一个句子,其开头首字母大写,如果不是句子则不大写】

%-------

% \subsection{\red{Web3 Based Money Laundering}}

% % 尽管区块链内在的透明度的透明度显示出追踪加密货币洗钱活动的前景，但对洗钱的测量和估计在加密货币中也很困难。加密货币洗钱的威胁不仅来自于不断膨胀的被盗资金数目，日新月异的洗钱手段也让监管者难以追上黑钱的脚步。加密货币生态中存在非法服务商为非法加密资产提供了洗钱服务，例如最近被美国OFAC制裁的去中心化混币服务商Tornado Cash。
% % 加密货币生态中还有一些服务提供商虽然并不普遍被认为是非法的，但也可能与犯罪活动有关或助推黑客进行洗钱活动，例如隐私币、赌博平台嵌套服务、主流交易所、以及基于智能合约的金融应用（如去中心交易所、DeFi）。
% While the inherent transparency of blockchain shows promise for tracking cryptocurrency’s money laundering activity, the measurement and estimation of money laundering is also difficult in cryptocurrencies. The threat of cryptocurrency money laundering comes not only from the ever-expanding number of stolen funds, but the ever-changing methods of laundering coins that make it difficult for regulators to keep up with the dirty money. In the cryptocurrency ecosystem, there are illegal service providers that provide money laundering services for illicit crypto assets, such as Tornado.Cash, a decentralized mixing service provider recently sanctioned by U.S. OFAC.
% There are also service providers in the cryptocurrency ecosystem that, while not generally considered illegal, may also be associated with criminal activity or facilitate money laundering by hackers, such as gambling platform, mainstream exchanges, and smart contract-based financial applications (like decentralized exchanges, DeFi etc.).

% %----------

% %\subsection{Account and Transaction}

\section{Related Work}

\subsection{Anti-Money Laundering Techniques}
\label{subsec:aml}
%---------------------修改版本---------------------------------
%在传统金融场景下，机器学习和深度学习的方法被用来反洗钱，which依赖身份关联信息，但匿名区块链系统很难获得这些数据。 加引用！
%在加密货币的世界中，介绍Elliptic。这个数据集被广泛关注并且运用到许多工作中
%但是Elliptic的缺点
%所以，一个能体现web3丰富度和完整度的洗钱数据集亟待被提出。

In traditional financial scenarios, AML techniques can obtain and analyze money laundering data through identity-linked information, as well as various modeling and learning approaches. However, in anonymous blockchain systems, the identity information and the association between accounts are usually not easily accessible. In the world of cryptocurrencies, the first publicly available dataset related to money laundering was the Elliptic dataset, classifying Bitcoin transactions into licit and illicit categories. The Elliptic dataset has attracted much attention and has been widely followed and used in a number of studies~\cite{alarab2020comparative,lorenz2020machine,Kolachala2021SoK,turner2020discerning}. 
% For example, ~\cite{lorenz2020machine} used unsupervised methods and Active Learning to detect illicit activity on Elliptic Dataset and  ~\cite{alarab2020comparative} found the ensemble learning outperformed all other methods using  features derived from Bitcoin transaction graph when spotting illicit transactions. 
However, the Elliptic dataset remains inappropriate for developing and validating AML techniques on Web 3 for two reasons. First, the Elliptic dataset only has binary labels for money laundering transactions and does not contain further details on the events and stages of money laundering; second, the bitcoin platform that this dataset focuses on has very different transaction behaviors than Web3 because it does not support smart contracts 
and decentralized applications.
%However, since Bitcoin does not support smart contracts, the Elliptic Dataset does not capture the new challenges to money laundering posed by the addition of digital assets and DeFi, and further does not adequately reflect money laundering in the increasingly diverse Web 3 ecosystem. 
Therefore, for AML on Web3, a dataset that represents diverse transactions and behaviors on Web3 is urgently needed to be proposed.
%Thus, a crypto money laundering dataset that reflects the richness of Web3 money laundering activities with high usability is urgent to be proposed.
% \subsection{\red{Transaction-based Analysis of Blockchain}}
% Alsuwailem \textit{et al.}~\cite{Alsuwailem2020} reviewed the existing literature concentrating on papers published between 2015 and 2020 to understand the state-of-the-art in AML systems, and demonstrated the observations using the four types: supervised learning, unsupervised learning, data sources, evaluation methods, implementation tools, sampling techniques, and study regions. The following are the key observations: The most widely used methods in AML systems from the supervised category are Decision Tree, Radom Forest, and SVM; neural networks are most popularly used in the unsupervised category; accuracy, Area Under the Curve (AUC), and precision are used for model evaluation; and the majority of the data used for research was customer and transaction data from banks.

\subsection{Financial Security Issues on Blockchain}
\label{subsec:datasets}
%---------------------修改版本---------------------------------
%区块链生态上出现很多安全问题，譬如xxx，xxx，xxx。
%现有很多异常识别的数据集，并提出了一系列的方法。加引用！
    %例如，针对钓鱼，xxx，针对庞氏骗局xxx。举几个例子。
%然而这些方法通常都是通过用户或者平台报告的标签来训练模型。【这句还是不写了，跟我们的贡献点关系不大，而且我们也依赖标签】
%根据xxx报告，很多这些安全事件都会通过洗钱再通过交易所等平台套现。而现有工作通常都在关注骗局的开始未深入挖掘他们背后的洗钱行为，无法完全理解整个骗局或安全事件的始末。因此本工作，就从以太坊上的heists和scams出发，来获得对应的洗钱流程，从而更好的实现对安全事件全貌的理解。

% 区块链生态系统中的安全问题比比皆是，如钓鱼诈骗、庞氏骗局、洗盘交易和DApp攻击等。[1, 79, 15, 20, 23, 34, 40, 41]. 现存在一些用于异常识别的数据集，并提出了一系列的方法来解决这些问题。例如，Chen等人[9]收集了Ponzi合约标签，并提出了Ponzi合同检测方法。Wu等人[40]提出了一种基于网络嵌入的钓鱼网站识别方法，并披露了一个钓鱼网站诈骗数据集。现有的工作通常集中在安全事件的开始，而没有深入挖掘其背后的洗钱行为。据报道~cite{Chainalysis2022}，许多安全事件发生之后都会进行洗钱，通过交易所等服务提供商来提现。因此，现有的研究不能完全了解安全事件的整个故事。
Security issues abound in the blockchain ecosystem, such as phishing scams, Ponzi schemes, wash trading and DApp attacks, etc.~\cite{risk2014Malte,Ponzi2018Chen,chen2019market,Chen2020Phishing,trans2vec2020Wu,Li2021TTAGN,gao2020tracking,victor2021detecting,Su2021attacks,Wu2021Mixing}. There exist several datasets for anomaly detection and a series of approaches have been put forward to solve these issues. For example, Chen \textit{et al.}~\cite{Ponzi2018Chen} 
collect Ponzi schemes labels\footnote{\url{https://www.kaggle.com/datasets/xblock/smart-ponzi-scheme-labels}} and propose a Ponzi contract detection approach.
%built a regression tree model based on account and code features to detect Ponzi schemes in the Ponzi contract dataset.
Wu \textit{et al.} ~\cite{trans2vec2020Wu} propose a network-embedding based method for phishing identification and disclose a phishing scam dataset\footnote{\url{http://xblock.pro/\#/dataset/6}}.
%However, those techniques usually depend on label information reported by users or platforms to train models. 
Existing efforts are usually focused on the beginning of the security incidents without digging deeper into the money laundering behind them. It has been reported~\cite{Chainalysis2022} that many security incidents are followed by money laundering to withdraw cash through service providers such as exchanges. As a result, existing research cannot fully understand the whole story of security incidents.

\section{Study Design \& Data Collection}
\label{sec:study_design}

%我们的研究旨在系统性地调查加密货币洗钱的行为特征和影响，从其洗钱源头、中介和出口，到洗钱交易形成的复杂网络及其带来的经济影响。为此，我们的研究由以下研究问题（RQs）驱动。

Our research aims to systematically investigate the characteristics of crypto money laundering in Web3, from an individual account, to the transaction networks formed by money laundering groups, and further, their resulting economic impact on the Web3 ecosystem. To this end, our research is driven by the following research questions (RQs):
\begin{enumerate}[RQ 1]
    \item \textit{\textbf{From a micro perspective, what are the characteristics of accounts and their transaction behaviors in the process of crypto money laundering in Web3?}}
    Previous work lacks the collection of data on crypto money laundering in Web3. We strive for a complete picture of cryptocurrency money laundering accounts that cover each suspicious path and compare their trading features with normal accounts. %\red{(what for???)}
    %and initially build the first open-source Ethereum money laundering dataset.

    \item \textit{\textbf{From a meso perspective, what are the properties of the complex network of transactions formed by crypto money laundering groups?}} As previous work has conducted network-based measurements and investigations on the entire Ethereum blockchain~\cite{Lee2020Measurements}, we wish to perform network modeling of money laundering groups' transactions to investigate the differences in money laundering networks compared to the entire transaction network. 
    
    \item \textit{\textbf{From a macro perspective, what is the impact of the flow of crypto money laundering on the economy of the Web3 ecosystem?}} Several reports~\cite{CipherTrace2022,Chainalysis2022,slowmist2022} have revealed that a large number of stolen assets have flowed into the Web3 trading platform. Therefore, it is interesting to explore how the inflow of stolen assets will affect the Web3 ecosystem. 
    
\end{enumerate}

% 1. 从微观的视角，加密货币洗钱过程中的账户及其交易行为有什么特点？以前的工作专注于案件事件本身和钱包黑名单地址，缺乏对非法资金清洗过程中的下游行为的系统性整理。本论文完成了对洗钱行为的完整刻画，覆盖每一条洗钱线路，初步建立以太坊反洗钱开源数据集。
% 2. 从中观的视角，加密货币洗钱团伙形成的交易复杂网络具有什么特点？由于之前的工作对以太坊整个区块链进行了基于网络的测量和调查，我们希望对洗钱团伙的交易进行网络建模，来调查洗钱网络相较于整体网络的差异。
% 3. 从宏观的视角，加密货币洗钱的流向对生态的经济方法造成什么样的影响？一些报告已经揭示了大量的被盗资产流入了加密货币交易平台。因此探索这些被盗资产的流入是否会影响加密货币是很有趣的。

% 加一个workflow示意图，风格参考Trade or Trick文章

% \red{\textbf{Overview of our approach.} The overall workflow of our money laundering study framework is shown in Figure~\ref{fig:framework}, which is made up of * major components. }

% \begin{figure}
%     \centering
%     \includegraphics[width=0.6\linewidth]{Fig_eg_framework.png}
%     \caption{\red{(!!!!!) The framework of investigating crypto money laundering in Web3.}}
%     \label{fig:framework}
% \end{figure}

\subsection{Abstraction Model for Money Laundering}
\label{subsec:abstraction model}

% 由于我们的目标是测量以太坊中被盗加密货币的洗钱过程，因此我们首先提出了一个洗钱过程的抽象模型。 
% 形式化地，一个案件的洗钱过程可以定义为(P,L,I)，其中P是洗钱的放置，L是分层的过程，I是聚合过程，对应着前文提到的洗钱三过程。
% 具体来说，小偷H把从受害者获得的非法资金放置在放置节点的地址集合L中，这部分地址是赃款的源头，可以通过查阅区块链浏览器、案件新闻、官方公告等途径来获取。
% 拿到P之后，黑客通过发起多笔以太币或ERC20代币的交易，将P里的钱逐层传递到了分层阶段的地址集L里，循环往复，混淆来源。最后，小偷H把案件赃款聚合到地址集I进行提现。通常情况下，I里的地址为和交易所、混币服务商、赌博等相关的地址。T代表洗钱过程中涉及的交易，包括外部、内部和ERC20代币交易。

Since our goal is to measure the money laundering process of stolen funds in Web3, we first propose an abstract model of the crypto money laundering process. Formally, the money laundering process of a heist can be defined as a four-tuple:  $\mathcal{(P,L,I,T)}$, where $\mathcal{P}$, $\mathcal{L}$ and $\mathcal{I}$ represent the address sets of \textit{placement}, \textit{layering}, and \textit{integration}, respectively (corresponding to the three phases of money laundering mentioned in Section~\ref{subsec:moneylaundering}). $\mathcal{T}$ is a transaction set which represents the involved transactions during the money laundering process, including external, internal and ERC20 token transactions.

\begin{figure}[htbp]
  \centering
  %\vskip -1.5ex
  \includegraphics[width=0.75\linewidth]{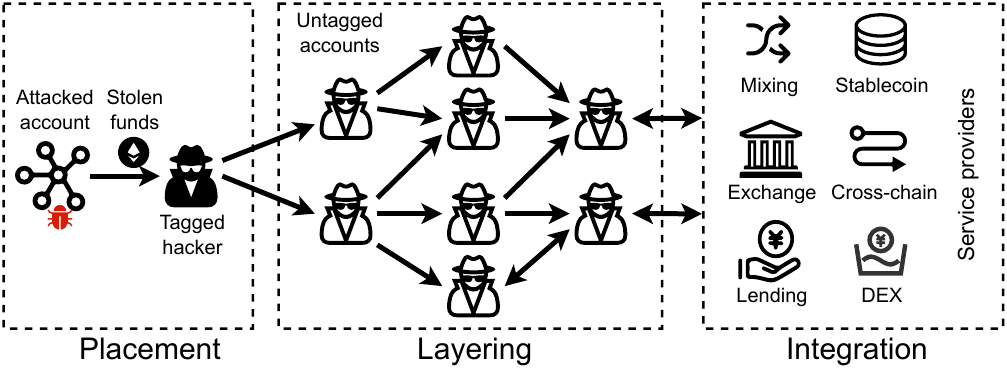}
  %\vskip -2ex
  \caption{Illustration of crypto money laundering phases in the Web3 ecosystem.}
  %\vskip -1.5ex 
  \label{fig:laundering_phase}
\end{figure}

% \mathcal{H} 没啥用，可以删了
% 图中显示了Web3中加密货币洗钱的一个toy example。首先，在placement阶段，黑客实施攻击，盗取资产并放在P集合的地址里。
Figure~\ref{fig:laundering_phase} shows a toy example of crypto money laundering in Web3. Specifically, the hacker performs an attack to steal assets and place them in $\mathcal{P}$, i.e. placement address set. The addresses in $\mathcal{P}$ is the source of the stolen funds, whose tagges can be obtained by consulting the blockchain browser (e.g. Etherscan) or official announcements. 
After taking $\mathcal{P}$, the hacker initiates multiple transactions of Ether or ERC20 tokens, passing the money in $\mathcal{P}$ layer by layer into the untagged layering address set $\mathcal{L}$ in the layering phase, cycling back and forth, obfuscating the source. Finally, the stolen funds are aggregated to integration address set $\mathcal{I}$ for cash out. The addresses in $\mathcal{I}$ are usually service providers such as exchanges, DeFi platforms, etc.

% mingxi's comment: 
% "the victims to the set of addresses $\mathcal{P}$", so $\mathcal{P}$ represents the set of addresses? As it should be the placement procedure; 
% "The addresses in $\mathcal{H}$ is the source of the stolen funds", the addresses in $\mathcal{L}$ or $\mathcal{I}$ is also the source of stolen funds?【确实没讲清楚。4.1部分准备重写】

\subsection{Framework of Dataset Construction}
%接着上述与洗钱过程的建模，我们在本小节将介绍洗钱数据集构建的基本方法，适用于本文分析的案例。主要环节包括：案件选择、数据爬取、洗钱交易追踪。
%Then, in this subsection, we present the basic framework of money laundering dataset construction, applicable to the incidents analyzed in this paper. The main phases include: target incident selection, data crawling, and transaction tracing.

\subsubsection{Target Incident Selection}

% 以太坊从创建至今，每年都发生了很多起加密资产盗窃案。截止至 2022 年 4 月 27 日，Etherscan上的Label Word Cloud 服务标记了115个地址为“Heist”，这些地址与交易所被黑、欺诈项目、DeFi攻击等被盗资产有关。
% 需要注意的是，这些数字只说明了来自 "加密货币原生 "犯罪的资金，其非法所得几乎总是以加密货币而非法币的形式获得。要调查线下犯罪所得的非法收入何时进入加密货币系统的（即放置阶段）是比较困难的。因此，本研究没有考虑贩毒或资助恐怖主义等线下犯罪活动。

% 根据Etherscan标记的Heist列表，我们按年份和被盗金额来选择了一些具有代表性的盗窃案件，并获取了每个案件在放置阶段的地址集。 到目前为止，被我们选择的案件列表在附录里展示了.

Ethereum has seen many security incidents of crypto asset theft each year since its creation. As of April 27, 2022, the ``Label Word Cloud'' service on Etherscan has flagged 115 addresses as ``Heist''\footnote{\url{https://etherscan.io/accounts/label/heist}} related to stolen assets from exchange hacks, scam projects, DeFi exploits, and more.
% Some of the accounts that stole ethereum crypto assets are flagged on Etherscan with the label "heist"\footnote{\url{https://etherscan.io/accounts/label/heist}}.
It should be noted that the statistics of Etherscan only account for incidents from ``cryptocurrency-native" crimes (i.e. on-chain crimes), in which illicit profits are almost always obtained in the form of cryptocurrency rather than fiat currency. 
% mingxi's comment: 大篇幅讨论off-chain crime很难，不在这篇paper的考虑范围内，好像有点问题。直接删去对paper好像不会造成什么影响，保留在这里反而觉得paper的贡献被削弱了。或者可以放在future work？实际上篇幅也不是很大，就一两句话，但感觉4.2.1也没几句话了【有道理，这句话我放到附录里面去讲论off-chain的事情】
Based on the ``Heist'' list marked by Etherscan, we selected a number of representative incidents by year and amount stolen, and obtained the placement address set  $\mathcal{P}$ for each incident.
The list of incidents selected by us for this study is shown in Table~\ref{tab:case_list} in Section~\ref{subsec:dataset_overview}. 
%mingxi's comment: ``Heist"? 建议看看pdf显示的结果是不是想要的显示结果，看起来有点怪【已修改为``xxx''来表示】

% \subsubsection{Data Crawling Tools}
% 放到了附录！

% zj comment: line6 的/beta在文中好像没有提到噢【确实】
\begin{algorithm}[t]
%\footnotesize
%\small
    \SetKwInOut{Input}{Input}
    \SetKwFunction{QuaryTxns}{QuaryTxns}
    \SetKwFunction{DirtyAmount}{DirtyAmount}
    \SetKwFunction{FilterTxns}{FilterTxns}
    \SetKwFunction{GetUnfamiliar}{GetUnfamiliar}
    \SetKwFunction{GetServices}{GetServices}
    \caption{Heuristic Transaction Tracing Algorithm}\label{algorithm}
    \label{alg:tracking}
    
    \KwData{placement address set $\mathcal{P}$, address label library $Lib$}
    \Input{max. depth of traced layers $K$, max. number of addresses per layer $\Psi$, threshold transaction number for unknown services $\Omega$}
    \KwResult{layering address set $\mathcal{L}$, integration address set $\mathcal{I}$, involved transaction set $\mathcal{T}$}
    
    % $\mathcal{L} \leftarrow \varnothing$\;
    % $\mathcal{I} \leftarrow \varnothing$\;
    % $\mathcal{T} \leftarrow \varnothing$\;
    $k \leftarrow 0$;   \tcp{The tracing depth} 
    $Cur_k \leftarrow \mathcal{P}$;  \tcp{The current suspicious address set}
    
    \While{$k\leq K$ and $0 < |Cur_k| < \Psi$}
    {
        \For{$a \in Cur_k$}
        {
            $\mathcal{T}_a \leftarrow $ \QuaryTxns{$a$}\;
            \If{\DirtyAmount($\mathcal{T}_a$, $\bigcup_{i=0}^{k} Cur_i$) > $\beta$}
            {
                \eIf{$|T_a|>\Omega $}
                {
                    $\mathcal{I} \leftarrow \mathcal{I} \cup $ $\{a\}$\;
                    $Cur_k \leftarrow Cur_k - \{a\}$\;
                    $T_a \leftarrow $ \FilterTxns($T_a$, $\bigcup_{i=0}^{k} Cur_i$)\;
                }
                {
                    $Cur_{k+1} \leftarrow  Cur_{k+1} \cup $ \GetUnfamiliar($\mathcal{T}_a$, $Lib$)\;
                    $\mathcal{I} \leftarrow \mathcal{I} \cup $ \GetServices($\mathcal{T}_a$, $Lib$)\;
                }
                $\mathcal{T} \leftarrow \mathcal{T} \cup \mathcal{T}_a $\;
            }
        }
        $k \leftarrow k+1 $\;
    }
    $\mathcal{L} \leftarrow \bigcup_{i=1}^{k} Cur_i$\;
\end{algorithm}

\subsubsection{\red{Tracing Method}}
\label{subsec:tracing method}
%为了构建上述的一个洗钱过程模型，我们需要有一定的策略从海量的区块链数据中进行交易追踪并提取用于洗钱的交易记录。
%为此，我们提出了一种基于启发式的算法来识别抢劫案的洗钱交易，如算法1所示. 算法的基本思想是，源头的放置地址集作为追踪的第0层地址，追踪每一笔转出交易的对象。后续未被标记的地址逐层按跳数进行分层，带标签的地址或具有其他显著交易特征的地址作为洗钱出口，即作为聚合地址集。
% 接下来 我们简要说明追踪算法的每个部分。

%\textbf{\red{[Add cross-chain tracing details in Algorithm 1. By Dan Lin ]}}
In order to build a model of the above-mentioned money laundering process of Web3 heists, we need to design strategies to trace transactions and sample the transactions used for money laundering.
To this end, we propose a heuristic-based algorithm to identify the money laundering transactions of the heists, as shown in Algorithm~\ref{alg:tracking}. 
The basic idea of the algorithm is the Augmented Poison Policy~\cite{risk2014Malte}. That is, the downstream accounts for money laundering are usually also money laundering accounts, only to the service provider as an exit.
% to take the placement address set as the Layer 0 for tracing, and keep tracing down the recipients of each out-going transaction. Unfamiliar out-going addresses that are not tagged are marked as the next layer of tracing and these addresses are also added to the layering address set. We trace layer after layer until service addresses or address with trading platform characteristics is encountered as the exit of the stolen funds, i.e., integration address set.

% 对于每个案件，我们已有标签库，且已知案件的放置阶段地址。该算法的输入是跟踪参数：跟踪层的最大深度$K$，每层的最大地址数$Psi$和未知服务的阈值交易数$Omega$. 设置这些参数的目的是为了控制交易追踪的范围，避免下游地址的数目爆炸，并且划定了终止追踪的条件。

% 交易追踪从放置地址集P开始（行1-2）.对于当前地址集的任意一个地址a，查询其交易记录（包括外部、内部、ERC20的交易）,得到交易记录Ta. 如果改地址的交易记录数目大于阈值，我们认为该地址是一个聚合阶段的未知服务地址，而非分层阶段的地址。这是因为我们认为洗钱的目的是掩人耳目，因此过程往往是十分低调的而不会使用一个地址进行大量的交易。因此，对于交易数量超过阈值（1000 甚至 10000）的账户，往往不是洗钱活动中分层的主力账户，而是作为将黑钱汇入交易平台的未知服务商。同时，我们还找出上游洗钱账户交易与该未知服务商的第一次转入交易时间和最后一次转入交易，过滤出一周以内的交易记录作为疑似洗钱交易加入到参与交易集合T当中。

% 由于每个大节点可能接收多个普通洗钱节点
% 以节点的下一跳节点的汇入，需要在交易汇总文件中找到与该大节点进行的第一
% 次转入交易时间戳（min_time）及最后一次转入交易时间戳（max_time），保留
% 大数量级节点在时间段 t（min_time ≤ t ≤ max_time + 604800（1 周的总秒数） )
% 内的转出交易

Next, we briefly explain each part of the tracing algorithm. The input of the algorithm includes the placement address set $\mathcal{P}$ for each incident, and the tracing parameters: the maximum depth of traced layers $K$, the maximum number of addresses per layer $\Psi$, and the threshold transaction number for unknown services $\Omega$. The purpose of setting these parameters is to control the scope of transaction tracing, to avoid an explosion in the number of downstream addresses, and to delineate the conditions for terminating tracing. Transaction tracking starts with the placement address set $\mathcal{P}$ (line 1--2). For any address $a$ in the current address set $Cur_k$ at layer $k$, query its external, internal and ERC20 transactions, and get the transaction record $\mathcal{T}_a$ (line 5). 
% mingxi's comment: 看到这里看明白了，所以$\mathcal{P}$是the placement address set. 感觉section4.1最开始那里好像说得不是特别清楚。那里说的是 "corresponding to the three phases of xxx"【有道理，重写4.1的部分】
We assume that the purpose of money laundering is to conceal the origin of illicit funds and thus the process tends to be very low profile and avoids using one address for a large number of transactions. Therefore, money laundering usually involves intensive and large-amount transactions between a group of accounts.
% money laundering is usually intensive???? and contains large-amount transactions.
% zj comment 上面这句话有点长，在第二个Therefore这里可以断一下
% ys comment 洗钱是大交易，这句话感觉可以改一下，应该是包含大交易【有道理，改成contains large-amount transactions】
We consider the address $a$ with a large number of transactions to be an unknown service address in the aggregation phase rather than the layering phase (lines 6--9), after filtering transactions containing small amounts of dirty money ($\leq$ threshold $\beta$).
%After filtering the transactions for small amounts of dirty money (threshold $\beta$), if the transaction number of address $a$ is great, we consider the address to be an unknown service address in the aggregation phase not in the layering phase (line 6--9). 
We then retain the transactions between unknown service providers and upstream laundering accounts $\bigcup_{1}^{k} Cur_i$ within one week as suspected money laundering transactions (line 10).
% zj comment 感觉下面这句话读起来比较困难:(，主要有两点：1. Otherwise是对应到上上句话而不是上句话，会有点割裂；【有道理，把otherwise换掉了】2. 这里对算法line 12 getUnfamiliar的解释太简单了("look for unmatched addresses based on the label library")，一下子看不太懂是怎么做的，感觉这个部分作为文章主要贡献之一，可以考虑增加一些篇幅，多一两句话讲清楚
%For the address $a$ with a small number of transactions, we look for unmatched addresses based on the label library as the suspicious addresses to be traced in the next layer, and add the matched service addresses to the aggregated address set (line 11--14). 
For address $a$ with a small number of transactions, we select the next level of suspicious addresses $Cur_{k+1}$ from recipient addresses of $a$'s outgoing transactions. In particular, we check whether these recipient addresses are known service providers, according the address label library $Lib$ %(The construction of $Lib$ is detailed in Appendix). 
If they are, they are added to $\mathcal{I}$. Otherwise, they are included in the $Cur_{k+1}$ crawled in the next layer (line 12-13).
% 对于有少量交易的地址a，我们从a的转出交易的接受方地址中找到下一层可疑地址。我们查找这些脏钱接受方的地址是否在服务商标签库中。若是，则加入到聚合层I，否则，列入下一层爬取的地址列表。
Then, the transactions of address $a$ are added to the transaction set $\mathcal{T}$ (line 15).
We keep increasing the depth $k$ (line 18) until the depth exceeds the maximum number of layers $K$, or the size of the current addresses set $Cur_k$ exceeds the range $[0,\Psi]$ (line 3). Finally, we merge the addresses of each layer to obtain the final layering address set $\mathcal{L}$ (line 20). 

% , which means that our data constitute a lower bound on the actual extent of money laundering transaction activity.
% 凯欣comment：several parameters总结一下可能比较好（tracing parameters，四个符号）
% 当追踪的深度超过最大层数K，或者当前层数的地址集为空集或者集合内地址数目超出阈值Y，则停止交易追踪。

% 伪造代币论文：由于我们的目标是测量以太坊中的伪造加密货币，我们既需要1）完整的ERC-20代币列表，用于检测伪造代币；也需要2）整个以太坊交易数据集，用于分析与伪造加密货币有关的交易。因此，我们利用Geth3，一个广泛使用的以太坊客户端来同步以太坊的账本。我们已经同步了所有的区块，直到2020年3月18日，总共有超过960万个区块。从区块中提取的数据包含外部交易、内部交易、合约信息和合约调用信息。然后我们得到所有智能合约的字节码和创建者信息。然后，我们分析字节码，以确定一个合同是否实现了ERC-20代币。基于这种方法，我们已经确定了超过176K个ERC-20代币以及它们的创建者信息。我们进一步从代码或Etherscan中获得这些代币的元数据（例如，网站、总供应量、持有人等）。为了方便分析，我们使用ElasticSearch来存储这些结构化的数据，这为我们的特征研究提供了一个查询界面。

% 需要 1）安全事件列表，用于作为洗钱的起点；也需要2）这些安全事件的下游洗钱过程，用于分析与伪造加密货币有关的交易。

\subsubsection{Data Crawling Tools}

% 爬虫是完成本文数据集收集工作的重要手段。尽管现有的数据分析工作采用同步以太坊区块数据的方法（如论文1），但这种方法并不适用于本研究，主要原因是以太坊区块数据是按照时间排序的（时间密集型数据），但本文需要对给定源头地址的交易记录进行BFS的爬取（空间密集型数据），从区块数据中查找某个地址的所以交易较为耗时。而以太坊官方浏览器的API提供了一个根据地址提取所有参与交易的功能。因此基于API的数据构建方法更合适。具体地，我们采用了基于Etherscan的API实现的开源爬虫工具包——BlockchainSpider。
% 为了在洗钱中判断非法资金的流向和清洗手段，我们还利用BlockchainSpider的Label Spiders来爬取了与交易所相关的标签地址（Exchange， Dex， Kyber， Uniswap， Bancor 等）、嵌套服务地址（OTC）、混币服务商（Tornado.Cash）以及其他与实际洗钱活动中出现的标签地址，建立了一个足够覆盖洗钱出口的地址比对库。
% 代币数据库参考 Zheng 等[40]2020 年发布的 ERC20TokenInfo与 ERC721TokenInfo
% 数据集，该数据集包括截止至第 13250000 个区块上发布的代币合约，包括合约地
% 址、代币名称、代币符号、市场规模等信息。该数据能够帮助识别代币交易中被用
% 于洗钱的代币种类，对代币的合规性进行初步判断
% 在本文的讨论中，我们将 Etherscan 上无标签标记的地址视为陌生地址。

% Although existing data analysis work uses the method of synchronizing the Ethereum block data (e.g., \cite{Lee2020Measurements}), this method is not applicable to this study. As discussed in \cite{Lin2021Evolution}, the reason is mainly because the block data is sorted by time (i.e., time-intensive data), but what we need is to crawl the transaction records of a given source address (i.e., space-intensive data). 
% Finding all transactions for a given address from the block data is much more time-consuming, whereas, the API of the official Ethereum browser provides a function to extract all the participating transactions based on the address. Therefore, the API-based data construction method is more appropriate. 
Crawlers are important means to accomplish the collection of the dataset in this work. 
Specifically, we used $\textsf{BlockchainSpider}$~\cite{wu2022transaction}, an open source crawler toolkit implemented based on the Etherscan API, to obtain the transaction records of accounts, i.e., the $\textsf{QuaryTxns}$ function in line 4, Algorithm~\ref{alg:tracking}. 
% 具体来说，我们采用BlockchainSpider，一个基于API的开源工具，来进行账户交易的爬取，即在tracing算法的QuaryTxns（line 4）部分。
% 在算法的line 12-13，我们还用到了地址标签库Lib，它由两部分构成，服务平台的标签和Token合约的标签。
Moreover, we utilize the address label library $Lib$ in line 12-13 of Algorithm~\ref{alg:tracking}. The label library $Lib$ consists of two parts: the labels of the service platforms and token contracts.
%对于服务商的标签，我们采用
To determine the service providers, we employ ``Label Spider'' of $\textsf{BlockchainSpider}$~\cite{wu2022transaction} to crawl label addresses associated with exchanges (e.g. ``Exchange'', ``DEX'', etc.), mixing services (e.g. ``Tornado.Cash'') and other label addresses that appear in connection with actual money laundering activities. We obtain more than 260,000 items, which is sufficient to cover money laundering destinations. 
The token contracts refer to the ``ERC20TokenInfo'' dataset with more than 313,000 ERC20 tokens, and the ``ERC721TokenInfo'' dataset with more than 15,000 ERC721 tokens, published by Zheng \textit{et. al.}~\cite{Zheng2020XBlockETH}, including contract addresses, token names, token symbols, etc. These can help identify the types of tokens being used for money laundering in token transactions. 
% In the discussion of this paper, we will treat addresses on Etherscan that are not tagged as \textit{unfamiliar} addresses.
% Theses datasets contain the token contracts issued on blocks up to the 13,250,000-th block
% 

\subsection{Example: Upbit Hack Case}

%We use a real hack called the Upbit Hack (mentioned in Section~\ref{subsec:datasets}), as the motivating example to show the challenge to trace the flow of money laundered based on cryptocurrency. %According to the official announcement of Upbit exchange, 342,000 ETH in Upbit's hot wallet was transferred to an unknown wallet address by hackers, and then the money was passed layer by layer.

% We show complexity of crypto money laundering with a case study of the Upbit hack labeled by Etherscan. Figure~\ref{fig:upbit_hack_example}(a) shows a simplified first half of money laundering process (without showing round-trip transactions) of Upbit exchange hack dataset\footnote{\url{https://etherscan.io/accounts/label/upbit-hack}} including the downstream money laundering addresses marked by Etherscan (813 addresses tagged as ''Upbit Hacker'', and 13,792 addresses tagged through transactions by an account named ''Upbit Savior'')

\begin{figure}
    \centering
    %\hspace{-2ex}
    \subfigure[]{
			%\centering
			\includegraphics[width=0.4\linewidth]{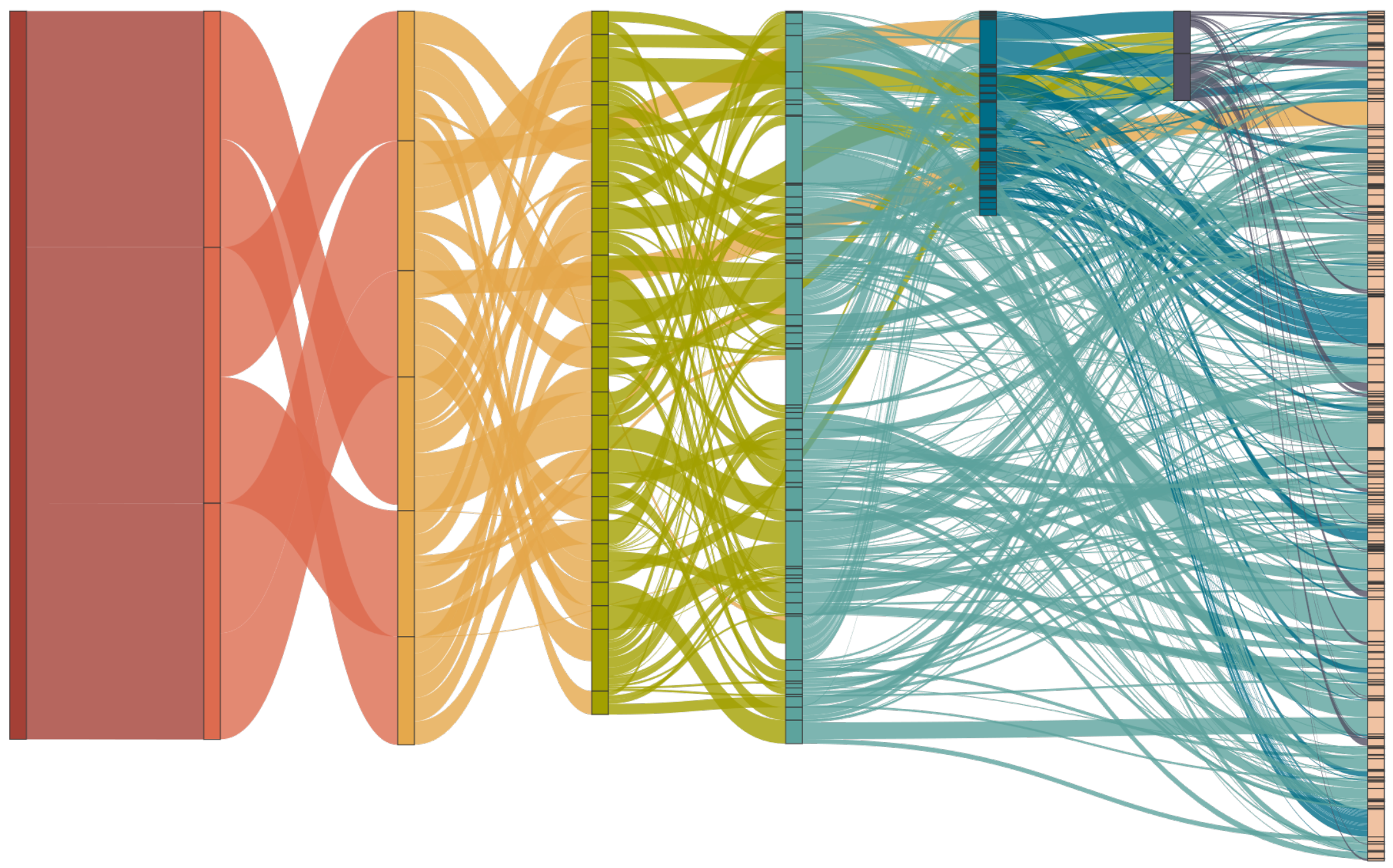}
	}
	%\hspace{-1ex}
	%\vspace{-1ex}
	\subfigure[]{
        \includegraphics[width=0.34\linewidth]{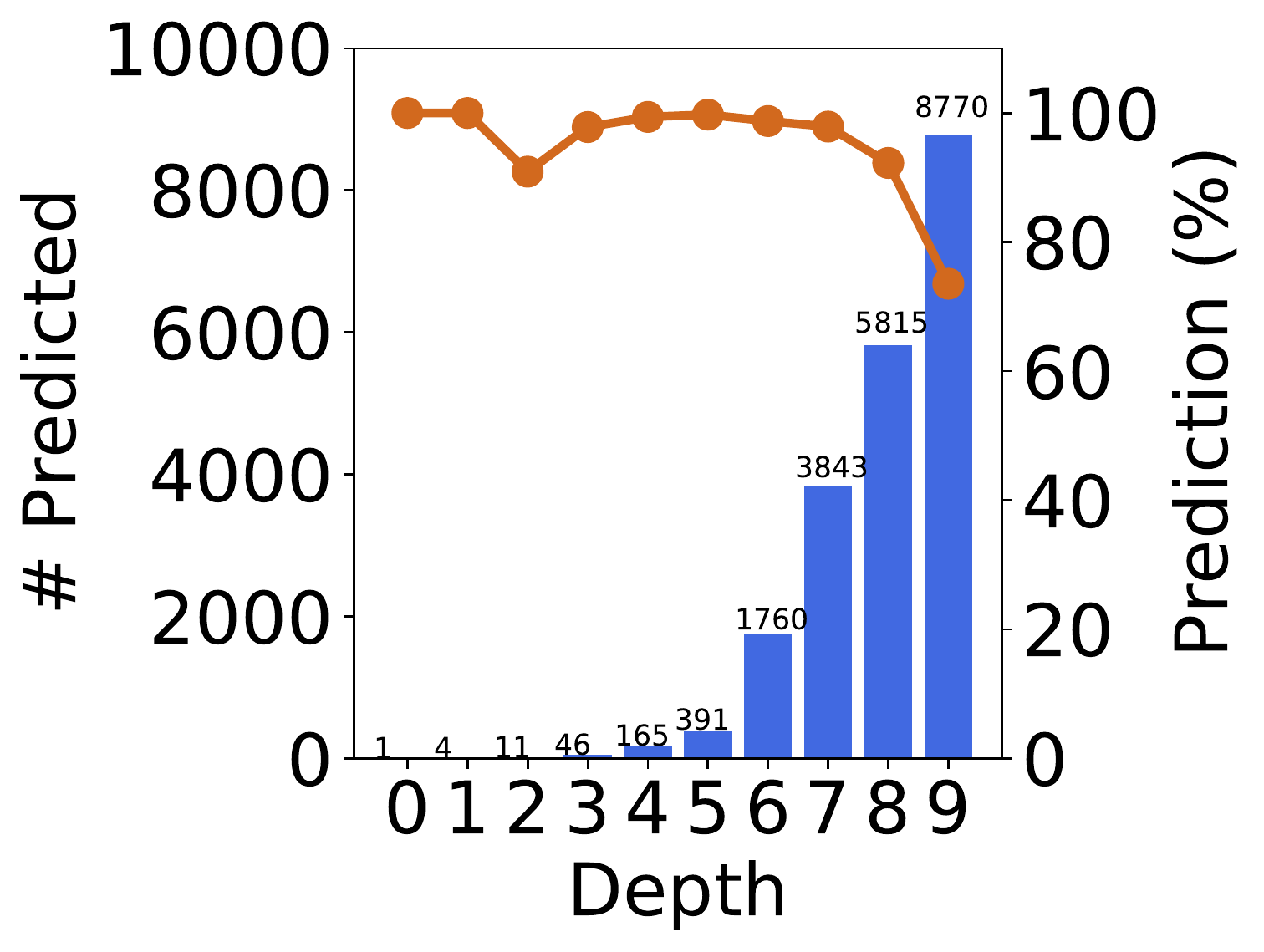}
    }
    
    %\vspace{-1ex}
    
    \caption{(a) The simplified money flow graph of the Upbit Hack case. Nodes represent accounts, and edges represent transactions. (b) Evaluation of Upbit Hack case.}
    % 橙色的折线是precision，蓝色的柱子是预测洗钱账户的数目
    \label{fig:upbit_hack_example}
    
    %\vskip -2ex
    
\end{figure}

To show the complexity of crypto money laundering, we visualize the simplified money flow graph of Upbit Hack (without round-trip transactions) in Figure~\ref{fig:upbit_hack_example}(a). Specifically, the root node (i.e., the leftmost node) represents the source of money laundering, (i.e., $\mathcal{P}$), the following nodes show the layering addresses (i.e., $\mathcal{L}$), and the links show the tainted money flow through multiple transactions (i.e., $\mathcal{T}$). Through this case, we can see that crypto money laundering flows are massively intertwined.

% 由于缺乏洗钱地址的ground truth，我们难以通过大规模的实验来证明我们方法的有效性。我们也并不能保证追踪的交易一定参与了洗钱。好在etherscan有一个唯一标记黑客洗钱账户的案例，也就是上面提到的Upbit 交易所盗窃案。

The lack of ground truth for money laundering addresses makes it difficult to show the effectiveness of our tracing method through large-scale experiments. %\red{It is not guaranteed that the traced transactions are really associated with money laundering.}
But the good thing is that Etherscan has a unique case of flagging a hacker's money laundering account, which is the Upbit Hack case\footnote{\url{https://etherscan.io/accounts/label/upbit-hack}} discussed here. 
% ys comment The good thing这一句句首可以加一个But逻辑转折，本段But之后表达的意思与But之前的一句形成转折。虽然缺少，但我们仍然可以，这段后面之后说的都是可以的原因和方法【有道理！】
%因此，我们用上述tracing方法得到疑似黑客的洗钱地址后，用etherscan上提供的标签（813个地址被标记为 ''Upbit Hacker'',和13792个地址通过交易被名为''Upbit Savior''的账户标记）计算precision指标，来验证我们追踪方法的有效性。
Therefore, we start from the source of Upbit Hack case and obtain the suspicious laundering addresses with Algorithm~\ref{alg:tracking} (Here we choose conservative parameters $K=20$, $\Psi=10,000$, $\beta=0.01$, and $\Omega=1,000$.) 

Then, as given in Figure~\ref{fig:upbit_hack_example}(b), we calculate the precision values with varying tracing depth to verify the effectiveness of our tracing method. 
% 图b展示了本文预测upbit hack 洗钱账户的precision随着追踪深度（depth）的增加的变化。
We observe that as the depth increases, the number of detected money laundering addresses grows exponentially. Even when the depth reaches 8, the precision is still over 90\%. This result suggests that our proposed tracing method is somewhat convincing.

%Due to the lack of ground truth, we manually investigated our own ground truth to support the measurement. Specifically, for each service, we first collected transactions according to the common features we observed from the sample transactions and then filtered false positives manually. Then we were able to evaluate the robustness of the proposed algorithm by comparing the result with the ground truth.

% 启发式方法的结果不错，是不是意味着直接用这个方法就好了呢？不，基于规则的方法适用于监管不到位的情况，而在有监管的情况下，比如设置了一些判断的基本规则，那么犯罪分子一定会想办法去规避。

\subsection{$\textsf{EthereumHeist}$ Dataset Overview}
\label{subsec:dataset_overview}
% Table generated by Excel2LaTeX from sheet '20221005'
\begin{table}%[htbp]
  \centering
  \caption{The selected incidents in Web3 from 2018-2022.}
    % 被选择的33个有代表性的Web3案件
  %\vskip -2ex
  \scalebox{1}{
    \begin{tabular}{lll|lll}
    \toprule
    \multicolumn{1}{l}{\textbf{Case Name}} & \textbf{Case Type} & \textbf{Year} & \textbf{Case Name} & \textbf{Case Type} & \multicolumn{1}{l}{\textbf{Year}} \\
    \midrule
    \midrule
    CoinrailHacker & CEX Hack & 2018  & LiquidExchangeHacker & CEX Hack & 2021 \\
    BancorHacker & DeFi Exploit & 2018  & AlphaHomoraV2Exploiter & DeFi Exploit & 2021 \\
    SpankChainHacker & Others & 2018  & bZxPrivKeyExploiter & DeFi Exploit & 2021 \\
    FakeMetadiumPresale & Scam  & 2018  & CreamFinanceExploiter & DeFi Exploit & 2021 \\
    BitpointHacker & CEX Hack & 2019  & EasyfiHacker & DeFi Exploit & 2021 \\
    CryptopiaHacker & CEX Hack & 2019  & PolyNetworkExploiter & DeFi Exploit & 2021 \\
    DragonExHacker & CEX Hack & 2019  & UraniumFinanceHacker & DeFi Exploit & 2021 \\
    UpbitHacker & CEX Hack & 2019  & BadgerDAOExploitFunder & Others & 2021 \\
    PlusTokenPonzi & Scam  & 2019  & VulcanForged & Others & 2021 \\
    KucoinHacker & CEX Hack & 2020  & ATOStolenFunds & CEX Hack & 2022 \\
    AkropolisHacker & DeFi Exploit & 2020  & LCXHacker & CEX Hack & 2022 \\
    HarvestFinanceExploiter & DeFi Exploit & 2020  & CashioAppExploiter & DeFi Exploit & 2022 \\
    Lendf.MeHacker & DeFi Exploit & 2020  & FloatProtocolFuseExploiter & DeFi Exploit & 2022 \\
    WarpFinanceHacker & DeFi Exploit & 2020  & DEGOandCocosExploiter & Others & 2022 \\
    NexusMutualHacker & Scam  & 2020  & Arthur0xWalletHacker & Scam  & 2022 \\
    AscendEXHacker & CEX Hack & 2021  & Fake\_Phishing5041 & Scam  & 2022 \\
    BitmartHacker & CEX Hack & 2021  &       &       &  \\
    \bottomrule
    \end{tabular}%
  }

  %\vskip -2ex
  \label{tab:case_list}%
\end{table}%

%%%%%%%%
% LinDan更新2022-10-4
%基于前文数据收集思路，我们从Etherscan的Hesit标签上选取了2018年至2022年四年间共33个Web3生态发生的代表性安全事件，按照上述框架模式进行数据收集
% ys comment 下面一段的Web3 ecology可改成Web3 ecosystem【已修改】
In this work, based on the above data collection method, we collect a total of 33 representative security incidents that occurred in the Web3 ecosystem from 2018 to 2022 based on Etherscan's ``Hesit'' tag. %The details of the modus operandi are shown in Appendix.
As shown in Table~\ref{tab:case_list}, there are four main types of Web3 cases collected in our dataset: CEX hack, DeFi exploits, Scams and Others, e.g. exploits of Decentralized Autonomous Organization (DAO), Game Finance (GameFi) and NFT Finance (NFTFi), etc.
% 从表1中可以看到，我们数据集收集的Web3案件主要有四个类型：CEX hack， DeFi exploits,  Scams和其他（包括exploit of GameFi，DAO,etc.）。CEX hack的主要作案手法是Security Breach，包括私钥泄露，恶意代码注入或者网络钓鱼攻击。DeFi exploits的主要原因包括两种：DeFi项目自身设计缺陷导致价格预言机被操纵，以及合约本身存在的代码漏洞如冲入漏洞。Scam的作案手法比较多样，包括钓鱼诈骗，旁氏骗局，虚假ico，还有最近出现NFT被骗事件。Others类型还包括近2年来较热门的GameFi和NFTFi。
We take a preliminary data analysis and exploration of the money laundering dataset as follows (The complete statistical table of case information is shown in Appendix):

\begin{enumerate}[(i)]

% 凯欣comment 有三个有趣的发现，案件的持续时间是按照什么来判断？
% 从案件的持续时间看，从天~1273天间不等，但一半地可以发现发生年份早的案件持续时间通常较长，例如Upbit Hack的持续时间2年之久，而持续时间短到一天之内的案件全部发生在2022年。这可能和洗钱手段的更新有关--混合服务。例如，2022年发生的LCX交易所事件，黑客只用了一天左右的时间，就通过去中心化交易所（DEX）将偷来的ERC20代币兑换成了以太币，最后全部转移到了名为Tornado.Cash的混合服务中。
\item In terms of duration, these cases range from less than 1 day to about 3 years. It can be found that some of the cases in early years usually last longer, e.g., the Upbit Hack laundering lasted for more than 2 years, while all the cases that lasted as short as one day occurred in 2022. This may be related to the newer means in Web3 - mixing service. For example, in the LCX exchange hack that occurred in 2022, the hacker took only about a day to exchange the stolen ERC20 tokens for Ether through a decentralized exchange (DEX), and eventually transferred them all to a mixing service named Tornado.Cash.
% 从涉案金额来看，这些案件的洗钱资金（ETH）平均在10的7次方数量级，最高可到达10的8次方数量级，可见案件造成的金额损失还是十分巨大的。
% ys comment 上面一段的newer means出现的时候后面跟了一个case，感觉可以先把means：混币说出来，再举case【已改~】

\item  In terms of the amount involved, the average amount of money laundered in these cases ranges from \$100 thousand to \$1 million, and the highest value reaches \$10 million, which shows that the financial loss due to the cases is still very huge.
% 另外，通过对各个案件的层数、gas花费以及交易数量，我们可以估计出案件的复杂度，总的来说，一个案件层数越大，涉及的中间帐户数以及交易数就越大，从而导致其案件数据庞大繁杂，黑客可以借此来掩盖自己的真实身份，但同时付出的代价是其gas花费也将相应提升，带来损失。

\item In terms of the complexity of the cases, we calculate the number of layers (i.e., tracing depth), transaction fees (i.e., gas cost), and the size of transaction set $\mathcal{T}$ of each case. In general, cases with more layers of money laundering have more accounts in $\mathcal{L}$ and transactions in $\mathcal{T}$, resulting in a larger and more complicated case data, which makes it easier for hackers to hide and conceal the source of stolen funds, but also costs hackers more in transaction fees.

\end{enumerate}

\section{RQ1: Trading Features of Accounts}

\label{sec:accounts}

For the laundering accounts of our dataset, we investigate their trading features such as transaction amounts, frequencies, lifespan, etc. To highlight the differences between layering and normal accounts, we also randomly sample the same number of normal accounts as reference objects. After comparing and observing the collected data, we obtain the most significant findings as follows.
% 对于我们数据集中L的洗钱账户，我们调查了他们的交易特征，如交易金额、频率、生命周期等。为了突出洗钱账户和正常账户之间的差异，我们还随机抽取了相同数目的正常账户作为参考对象，在观察和比较收集到的数据后，我们得到最重要的观察结果如下。
% 140 thousand normal accounts (of the same order of magnitude)

%修改：蕴湄1009

%1009修改前:, and obtain the following observations.

% zj comment: fig.3的lifespan缺少单位，以及不太看得懂纵坐标是什么意思，导致不太清楚图是什么意思【好的】
% 一方面，许多洗钱账户的寿命极短，表现出 "用完即弃 "的特点。
%\textit{Lifespan.} 
\subsection{Lifespan}
\begin{figure}
    \centering
    \subfigure[Lifespan distribution]{
			%\centering
% 			\includegraphics[width=0.37\linewidth]{Fig_Lifespan_1011.png}
			\includegraphics[width=0.3\linewidth]{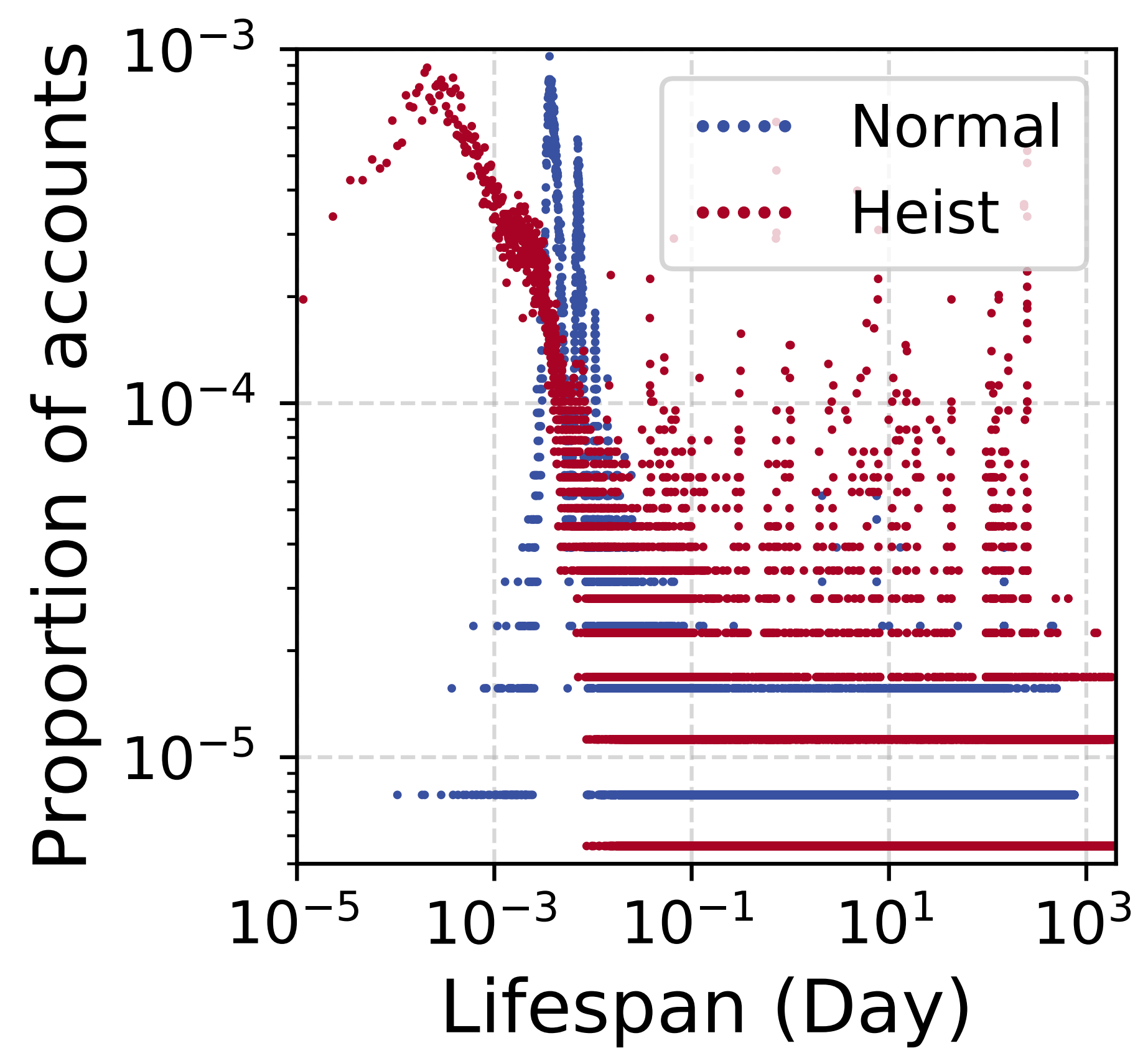}
	}
    \subfigure[Degree distribution]{
			%\centering
			\includegraphics[width=0.3\linewidth]{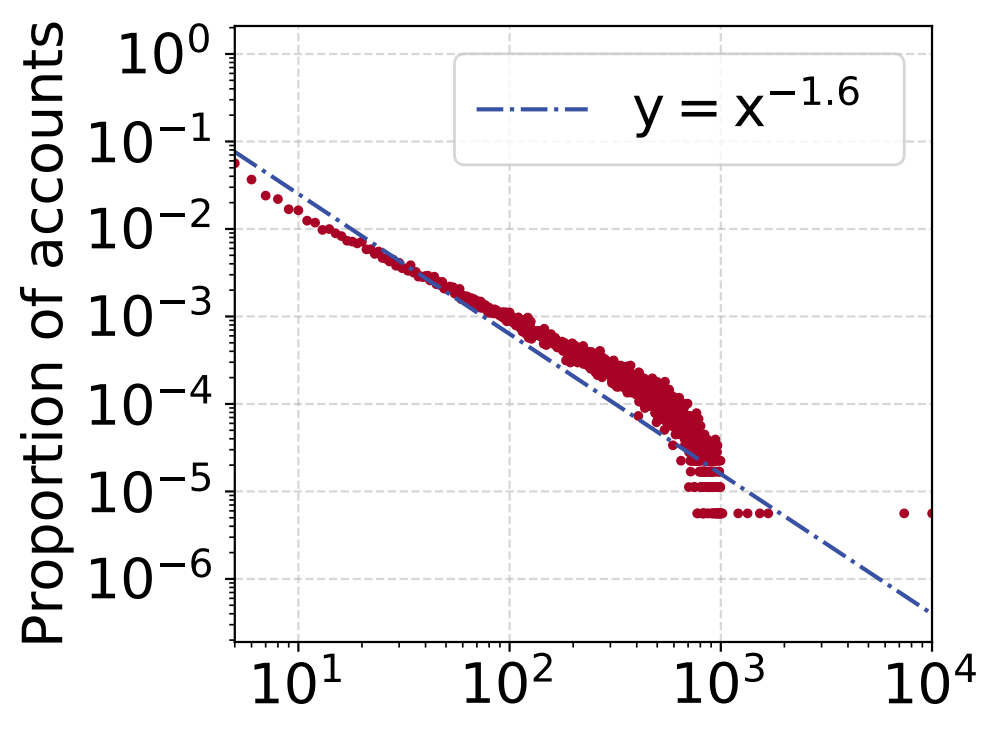}
	}
    %\vskip 1ex
	%%\hspace{-1ex}
	%\vspace{-1ex}
	\subfigure[Frequency distribution]{
        \includegraphics[width=0.3\linewidth]{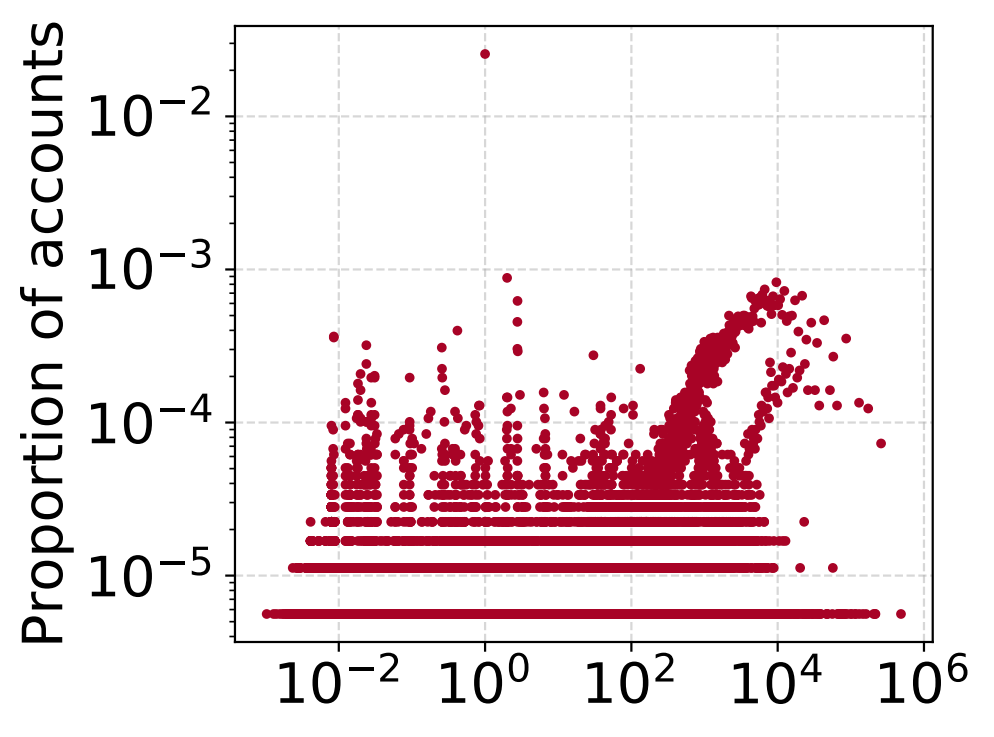}
	}
    %\vspace{1ex}
    % 这个图太小了！！！！
    %\hspace{-2ex}
    %\vspace{-2ex}
    %\caption{Trading characteristics of crypto money laundering accounts vs normal accounts.}
    \caption{Comparison of trading characteristics.}
    %\vskip -2ex

    % 橙色的折线是precision，蓝色的柱子是预测洗钱账户的数目
    \label{fig:account_feature}
\end{figure}

As shown in Figure~\ref{fig:account_feature}(a), on the one hand, many money laundering accounts have extremely short lifespans, exhibiting a ``used-and-dumped'' characteristic. Compared to normal accounts, the peak of the lifespan distribution for the money laundering account is more to the left.
%Figure~\ref{fig:account_feature}(a) shows that the peak of money laundering accounts is more to the left compared to normal accounts and their peaks are more rounded. 
On the other hand, money laundering accounts with larger lifespans show an irregularly high percentage of jumps, which is because some of the more careful hackers do not transfer stolen funds immediately, but lurk until the wind passes before laundering. For example, the Coinrail hacker\footnote{\url{https://cn.etherscan.com/address/0xf6884686a999f5ae6c1af03db92bab9c6d7dc8de}} stole assets in 2018 and then lurked for two years until 2020 when the stolen money was transferred out.

% 用之即弃、迟疑观望

%------------------------------

%针对数据集中所有被我们标记为洗钱的帐户，我们对其进行了账户特征的统计分析，通过统计其交易特征，包括交易次数、交易金额、交易频率以及洗钱账户的生命周期，同时随机抽样了与数据集中洗钱帐户数（178170）数量级相当的140560个正常帐户，同样对上述特征进行统计作为分析参照物。
% 最后总结本小节的发现

%\textcolor{red}{xxxx xxxx xxxx xxx xxx xxxxxxx x xxxxx xxxxx xxx xxxxxx xxxxxxxxx xxxx xxxx xxxx xxxx xxx xxx xxxxxxx x xxxxx xxxxx xxx xxxxxx xxxxxxxxx xxxx xxxx xxxx xxxx xxx xxx xxxxxxx x xxxxx xxxxx xxx xxxxxx xxxxxxxxx xxxx xxxx xxxx xxxx xxx xxx xxxxxxx x xxxxx xxxxx xxx xxxxxx xxxxxxxxx xxxx xxxx xxxx xxxx xxx xxx xxxxxxx x xxxxx xxxxx xxx xxxxxx xxxxxxxxx xxxx xxxx xxxx xxxx xxx xxx xxxxxxx x xxxxx xxxxx xxx xxxxxx xxxxxxxxx xxxx xxxx xxxx xxxx xxx xxx xxxxxxx x xxxxx xxxxx xxx xxxxxx xxxxxxxxx xxxx xxxx xxxx xxxx xxx xxx xxxxxxx x xxxxx xxxxx xxx xxxxxx xxxxxxxxx xxxx xxxx xxxx xxxx xxx xxx xxxxxxx x xxxxx xxxxx xxx xxxxxx xxxxxxxxx xxxx xxxx xxxx xxxx xxx xxx xxxxxxx x xxxxx xxxxx xxx xxxxxx xxxxxxxxx xxxx xxxx xxxx xxxx xxx xxx xxxxxxx x xxxxx xxxxx xxx xxxxxx xxxxxxxxx xxxx}

\subsection{Degree and Frequency}

\begin{figure}
    \centering
    
    %\vspace{-1ex}

        %\includegraphics[width=0.4\linewidth]{Fig_V_1009_old.pdf}
        % \includegraphics[width=0.58\linewidth]{Fig_V_1011_new.png}
        \includegraphics[width=0.5\linewidth]{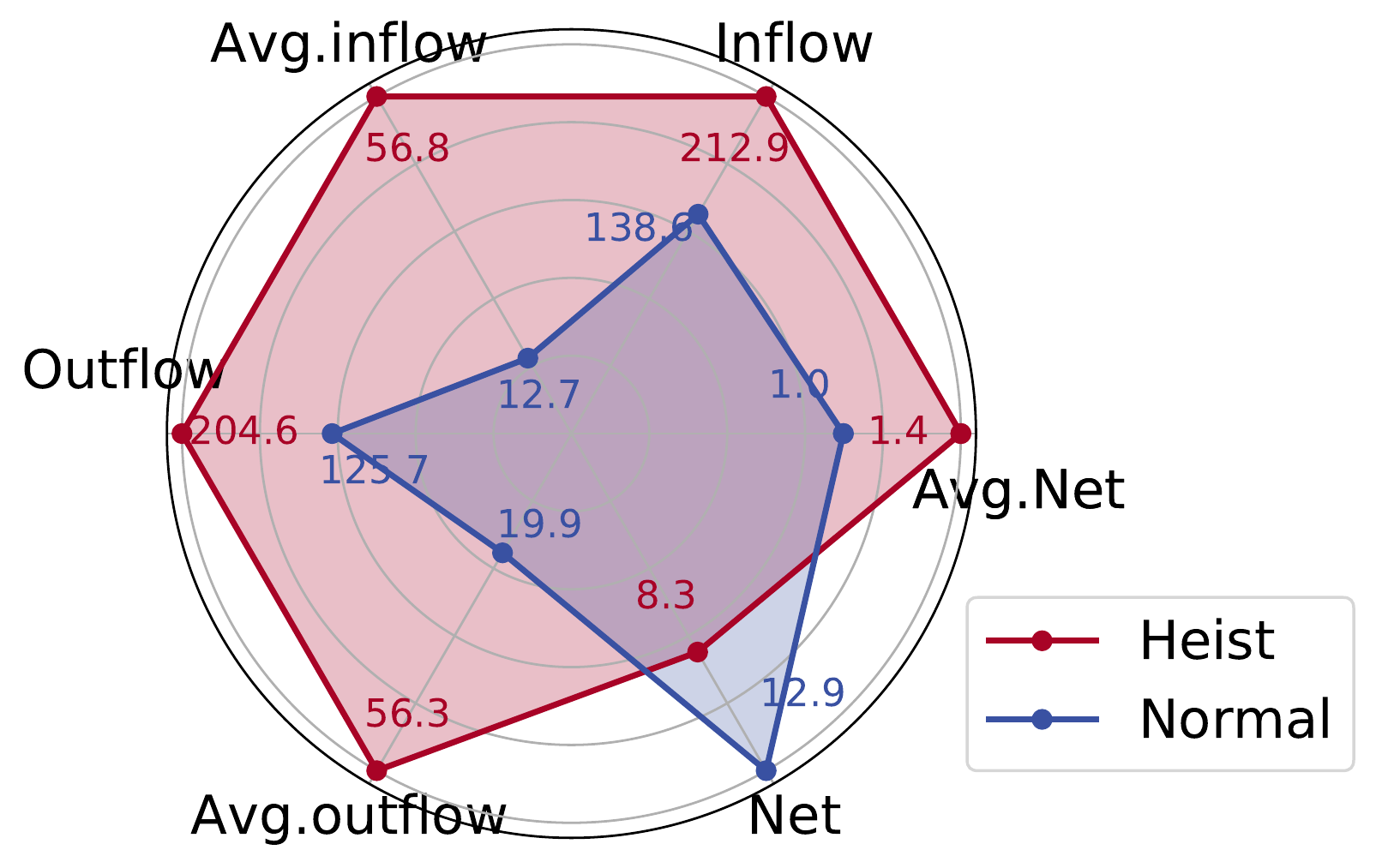}
        %\caption{transaction volumn}
    %1006：未更新旧图
    \caption{Comparison of transaction amount (ETH).}
    %\caption{(a) All-degree of Heist.  (b) Transaction frequency of Heist.}
    % 橙色的折线是precision，蓝色的柱子是预测洗钱账户的数目
    \label{fig:degree and frequency}
    
    %\vskip -2ex
    
\end{figure}

%\noindent $\bullet$ 
Degree indicating the transaction activeness, i.e., the sum of in-degree and out-degree. It can be concluded from Figure \ref{fig:degree and frequency}(a) that the net transaction follows power law distribution, we also plot the fitted line $y=x^{-\alpha}$ ($\alpha =1.6$) to prove this. Due to the threshold we set is $10^{3}$, we can see from \red{the} figure that our data cut at the same point. As for the several points\red{,} that fall around $10^{4}$ is because they get a big degree in the placement phase the very first time and escape from our threshold control. 
% %-------qs comment：不懂-----------------
%\noindent $\bullet$ 
Frequency is the number of \red{transactions} that \red{an account is involved} in per day. Generally\red{,} the frequency of a Heist account is evenly distributed over all magnitudes, but when it comes to the account that \red{has} a high proportion, we can see they normally have a high transaction frequency.
    
% 交易量。我们计算并统计了每个案件中洗钱账户的流入、流出金额和总金额以及账户对应的平均涉案金额，如图x所示。我们发现洗钱账户所涉及的交易额在每一个指标上都会明显大于正常帐户。这符合以往研究中对洗钱特征的假设：a dense flow of money transfers high volumes of money
%\textit{Transaction amount.} 
\subsection{Transaction Amount}
We calculate and count the inflow, outflow, and net value of layering accounts in each heist, as well as the corresponding average value per transaction of accounts, as shown in Figure~\ref{fig:account_feature}(b). Note that we filter the account whose transaction value is larger than 1,000 to reduce the bias caused by exchanges with frequent transactions~\cite{Chen2020Phishing}. We find that the transaction amount of layering accounts is significantly larger than normal accounts in almost every indicator. This shows that even though hackers can create accounts without restrictions, the amount of stolen funds is so large that the amount per transaction in the laundering process remains large. As shown in Figure~\ref{fig:account_feature}(b), the average inflow and outflow value of layering accounts reach over 50 ETH, which is about 3-5 times higher than that of normal accounts. % 50ETH per transaction
The reason for the smaller net value of laundering may be that as little as possible money is left  hackers usually leave as little money in the layering account as possible to reduce the risk of being frozen.
% 这说明尽管黑客可以不受限制地创建账户，但由于盗取资产的金额实在太大，使得每个洗钱账户的交易金额仍然很大，其输入和输出的数值平均高达约200ETH。
% This is consistent with the hypothesis of previous studies~\cite{flowscope2020,Ji2022Cash}. 

%考虑到交易所的交易量本身特别大，因此为了更准确刻画普通账户特征，我们过滤掉交易量＞1万的账户 引用:Phishing Scam Detection on Ethereum: Towards Financial Security for Blockchain Ecosystem
% dense subgraph
% on the characteristics of money laundering - a dense flow of money transfers high volumes of money

%%%%%%%%%%%%%%%%%%%%%%%%%%%%%%%%%%%%%%%%%%%%%%%%%%%%%%%%%%%
% WWW被注释掉的附录
% 在Section xx中，我们讨论了洗钱账户的交易特征，这里我们补充其他特征的分布，并给出其他发现。

%In Section~\ref{sec:accounts}, we have discussed the trading characteristics of money laundering accounts, here we present more characterization and give additional findings, as shown in Figure \ref{fig:degree and frequency}.
% we display two other characteristics of money laundering accounts through presenting the distribution of these features.

\vspace{1.5ex}
\begin{mdframed}
    \textbf{Finding 1.} The laundering accounts usually present an extremely short or long lifespan, reflecting the ``use-and-dump'' or ``wait-and-see'' strategies, respectively. Hackers tend to engage in higher value transactions, but their net transaction amount is smaller. The value flow in and out heist account reaches around 200 ETH on average.
\end{mdframed}
\vspace{1.5ex}
%%%%%%%%%%%%%%%%%%%%%%%%%%%%%%%%%%%%

% \noindent $\bullet$ \textbf{Finding 1.} 
% %总的来说，洗钱账户的生命周期呈现为极短或极长两种形态，且其在交易中通常会涉及到数额较大的交易。
% %In short, we concluded two main characteristics when it referred to money laundering 
% The laundering accounts usually present an extremely short or long lifespan, reflecting the ``use-and-dump'' or ``wait-and-see'' strategies, respectively. Hackers tend to engage in higher value transactions, but their net transaction amount is smaller.
% % with incoming and outgoing values averaging up to about 200 ETH.
% %We concluded two main characteristics when it referred to money laundering accounts. Firstly, They normally present an extremely short or long lifespan. Besides, they tend to involve in the transactions that require a high value of money, the value flow in and out heist account reaches around 200 ETH on average.
% % ys comment 感觉师姐可以给一个数量会更好，其实这一点也算是一个与传统的不同【有道理】

% \vspace{-1ex}

\section{RQ2: Network Features of Groups}
\label{sec:network_properties}

%\red{Add more introduction...}
% 在上一个section，我们介绍了孤立的洗钱账户和普通账户的特征异同，那么自然地我们会有下一个问题，这些洗钱交易网络和普通交易网络又有什么异同点呢？
In the previous section, we describe the similarities and differences in the characteristics of isolated money laundering accounts and normal accounts, so naturally we have the next question, what are the similarities and differences between these money laundering transaction networks and normal transaction networks?
%Writer: Yiyue Cao
%如果犯罪团伙以洗钱为目的发起交易，那么这些非正常交易构成的集合的特征可能与区块链上的总体正常交易集合的特征有差异。本section讨论洗钱交易集合的特征。为此，我们首先将每个洗钱案例构建为洗钱交易网络，and extract several features representing global properties of a network———also known as“summery features”. 然后我们从不同案件的洗钱网络中提取并统计了各类模体。 同时，我们将这些网络特征与以太坊上的总体正常交易网络进行了对比分析。
%If a criminal group initiates transactions for the purpose of money laundering, the characteristics of the set of irregular transactions may differ from the overall set of normal transactions on the blockchain. This section discusses the characteristics of money-laundering transaction group from a perspective of complex network. To this end, we first construct each money laundering case as a transaction network, and extract several features representing the global properties of a network — also known as "summery features". Then we pull and count the various high-order motifs from the money laundering networks in the different cases. At the same time, we compare these characteristics with the overall normal transaction network on Ethereum.

\subsection{Network Modeling}

% zj comment: "differ from the entire tx network", entire 是否换成normal/benign之类的词? 因为洗钱交易也是entire 交易网络的一部分【有道理，不过引用的论文确实是entire network，而我们的洗钱交易数据在entire network中的占比可忽略不计，这个在后面提了一下）】
If criminal groups initiate transactions for the purpose of money laundering, then these transaction networks may differ from the normal transaction network of Ethereum. To this end, we first model the money laundering transaction of each case as a network, $G= (V, E)$, $E$ is the edge set containing all transactions in the case, i.e. $\mathcal{T}$, and $V$ is a node set which denotes the accounts involved in these transactions.

% \red{as shown in Table~\ref{tab:diameter} in the Appendix, which is in line with our data collection strategy.}

% !!!!! 我们的案例的网络直径从xx到xx。在附录中有

%G Definition and Construction. 我们将每一个案例的洗钱交易过程建模为一个Network，G=（V，E），E是一个包含洗钱案例中所有交易的边集，V是一个交易中涉及的账户所被建模成的节点集。
%G Definition and Construction.We model the money-laundering transaction process of each case as a Network, G= (V, E), E is an edge set containing all transactions in the case, V is a node set modeled by the accounts involved in these transactions.

%网络直径是网络中任意两节点间的最长距离，我们首先通过计算网络直径来初步了解不同案件的洗钱网络规模。我们看到洗钱网络的直径和数据集的level成正比，见表X（or图），这符合我们的数据集收集策略。
% Network diameter is the longest distance between any two nodes in the network. 

%\vspace{-1.5ex}
\subsection{Global Network Properties}

%我们将案例根据交易货币的类型划分为ETH交易网络（包含外部交易、内部交易）和ERC 20代币交易网络，以便分别与以太坊的TransactionNet与TokenNet做对比。同时，我们也在不同案例之间进行比较，以探究Web3犯罪的洗钱过程是否可能与案例自身有关。
For each money laundering case, we model two networks: the Ether transactions (including external and internal transactions) as $\textsf{HeistEthNet}$, and the ERC20 token transactions as $\textsf{HeistTokenNet}$, in order to compare with $\textsf{TransactionNet}$~\cite{Lee2020Measurements} and $\textsf{TokenNet}$~\cite{Lee2020Measurements} of the entire Ethereum, respectively.  
% 值得注意的是，entire network可以反映normal交易网络的性质，因为我们的洗钱账户（16万）在整个网络的账户（4600万）仅占0.3%，可以忽略不计。
It is worth noting that the entire network~\cite{Lee2020Measurements} may reflect the nature of the normal transaction network, as our money laundering accounts represent a minuscule 0.3\% of the entire transaction network (46 million).
Comparative results of graph properties are shown in Table~\ref{tab:network_properties}.  %Also, we compare between cases to explore whether the money laundering process of Web3 crime may be related to the cases themselves.

%------------------------------------
%Writer: Yiyue Cao
%另外，每个案例可以根据交易货币的类型划分为ERC 20代币交易网络和ETH交易网络（包含外部交易、内部交易），以分别与以太坊上的正常网络[]进行对比：ERC20代币交易网络与TokenNet做对比，ETH交易网络与TransactionNet做对比。同时，我们也在不同案例之间进行比较，以探究Web3犯罪的洗钱过程是否可能与案例自身条件，如持续时间，类型等有关。
%In addition, each case can be divided into an ERC20 token transaction network and an ETH transaction network (including external and internal transactions) to compare with the normal network on Ethereum ~\cite{Lee2020Measurements}: the ERC20 token transaction network is compared with TokenNet, and the ETH transaction network is compared with TransactionNet. At the same time, we also compare different cases to explore whether the money-laundering process of Web3 crimes may be related to the conditions of the cases themselves, such as duration and type.

%We study several summary features of the networks. 它们从反映了洗钱网络的某些特性，尤其是与总体交易网络的差别。
%We study several summary features of the networks.They reflect certain characteristics of the money-laundering network, especially the difference from the overall transaction network.

\subsubsection{Basic Features}

% 我们统计了每个洗钱网络中的自环数量在所有边数中所占比例并计算其统计量，与正常网络进行对比（表Y）。与TokenNet对比时，和预期的一样，洗钱网络的自环占比更小。因为表现为自环的交易不符合洗钱的目的，即没有进行资产的分割和转移。惊讶的是，HeistEthNet的自环占比平均值高于TransactionNet。通过进一步分析发现，是因为CashioApp Exploiter案件的黑客通过大量自交易在input data区留下message给社区传递信息，导致这个案件的总体洗钱网络的自环占比达到了8.62%。总之，一般来说洗钱网络的自环占比低于正常网络。
First, we count the \textit{self-loop} ratio of each money laundering network and calculate their statistic. When compared to $\textsf{TokenNet}$, as expected, the self-loop ratio of $\textsf{HeistTokenNet}$ is smaller because self-loop transactions are not consistent with the purpose of money laundering, i.e., no splitting and diverting.
Surprisingly, the average self-loop ratio of $\textsf{HeistEthNet}$ is higher than that of $\textsf{TransactionNet}$. Our further analysis reveals that it is because the CashioApp Exploiter\footnote{\url{https://etherscan.io/address/0x86766247ba3405c5f15f06b895294200809e9cfb}} left messages to the community through several self-transactions in the input data area, resulting in the high self-loop ratio in this case. 
% 其次，我们发现，洗钱网络的互惠性高于整个网络的互惠性，这很可能与Web3洗钱过程中代币互换的高活跃度有关。

\textit{Reciprocity} is defined as the ratio of the number of edges pointing in both directions to the total number of edges.
We find that the reciprocity of money laundering networks is higher than that of the entire network, which is likely related to the high activity of token swaps in the Web3 money laundering process.
% 第三，
% 我们发现洗钱网络的图密度均值显著高于全网络，这意味着洗钱网络是一个频繁密集的子网络。更多实验结果见附录。
% there are a lot of multiple arcs between vertices in the MultiDigraph (all directed arcs between a pair of vertices are retained) of ContractNet.
% 最后，我们报告了网络的密度，按照现有研究[1]中的计算公式。

Finally, we report the \textit{density} of networks, following the formulas in the exiting research~\cite{Lee2020Measurements}.
% 我们可以看到，HeistEthNet和HeistTokenNet在multidigraph中的平均density都比他们在simple undigraph的两倍还多，其中HeistEthNet甚至达到了4倍。但是在entire 网络中，transaction net在multidigraph的密度只是他在simple undigraph的1.5倍。这意味着洗钱网络是一个频繁密集的子网络。
As shown in Table~\ref{tab:network_properties}, both $\textsf{HeistEthNet}$ and $\textsf{HeistTokenNet}$ have more than twice the average density in multidigraph than they have in simple undigraph, with $\textsf{HeistEthNet}$ even reaching 4 times. But in the entire network, the density of $\textsf{TransactionNet}$ in multidigraph is only 1.5 times of that in simple undigraph. This indicates crypto money laundering networks are frequent and dense sub-networks.
\begin{table}%[htbp]
  \centering
  %\vspace{-1ex}
  %\caption{Network properties of the heist laundering networks and the entire Ethereum network (``Med.'' means median. ``Avg.'' means average).}
  \caption{Comparison of network properties. (``Med.'' means median. ``Avg.'' means average).} 
  %\vspace{-2ex}
  \scalebox{1}{
    %\hspace{-2ex}
    \begin{tabular}{lcccccc}
    \toprule
          & \multicolumn{1}{c}{\textbf{Self-loop}}
          & \multicolumn{1}{c}{\textbf{Reciprocity}} 
          & \multicolumn{1}{p{3em}}{\textbf{Density\newline{}(s.,undi)\tablefootnote{simple, undirected graph}}} 
          & \multicolumn{1}{p{2.9em}}{\textbf{Density\newline{}(multidi)}}  
          & \multicolumn{1}{p{2.9em}}{\textbf{Global\newline{}cluster}}
          & \multicolumn{1}{p{2.8em}}{\textbf{Avg.\newline{}pathlen}}  \\
    \midrule
    \midrule
    $\textsf{HeistEthNet}$ (Med.) & 0.06\% & 4.62E-02 & 2.59E-02 & 2.67E-02 & 1.37E-02 & 2.47 \\
    $\textsf{HeistEthNet}$ (Avg.) & 0.64\% & 9.78E-02 & \textbf{1.86E-01} & \textbf{7.70E-01} & 2.71E-02 & 5.56 \\
    $\textsf{TransactionNet}$~\cite{Lee2020Measurements} & 0.13\% & 3.00E-02 & 1.24E-07 & 1.87E-07 & 1.00E-01 & 5.33 \\
    \midrule
    $\textsf{HeistTokenNet}$ (Med.) & 0.02\% & 4.21E-02 & 2.83E-02 & 3.36E-02 & 7.08E-03 & 2.44 \\
    $\textsf{HeistTokenNet}$ (Avg.) & 0.03\% & 1.01E-01 & 1.78E-01 & 3.80E-01 & 1.15E-02 & 4.97 \\
    $\textsf{TokenNet}$~\cite{Lee2020Measurements} & 0.19\% & 3.00E-02 & 2.03E-07 & 1.87E-07 &  1.75E-01 & 3.87 \\
    \bottomrule
    \end{tabular}%
  }
  %\vspace{-2ex}
  \label{tab:network_properties}%
\end{table}%

\subsubsection{Small-world Behaviour}
%\subsubsection{Advanced analysis of small-world network}
% %Writer: Yiyue Cao
%绝大部分真实网络的平均距离都短的超乎我们的想象，人们将网络规模很大但是平均距离很小的性质形象地称为小世界效应。Analogous to social networks, blockchain graphs are also small-world.[] 在表Y中，我们发现，洗钱网络的average shortest pathlength 和区块链正常交易网络一样处于4~6。然而，通过计算每一个案例对应的洗钱网络的平均聚类系数，我们发现洗钱网络的聚类系数比正常网络小（包括ERC20代币交易网络和ETH交易网络）。 这可能是由于洗钱的特殊目的使其丧失了正常网络的多枢纽和社交性的特征。所以我们发现，虽然洗钱网络的average shortest pathlength足够小， 但是其聚类系数不大，因此洗钱网络并不具备小世界效应。
%The average distance of most real networks is short beyond our imagination. 
Researchers refer to the property of large network size but small average distance as small world effect. Analogous to social networks, the entire Ethereum blockchain graphs are also small-world~\cite{Lee2020Measurements}. In Table~\ref{tab:network_properties}, we find that the average shortest \textit{path length} of $\textsf{HeistEthNet}$ and $\textsf{HeistTokenNet}$ is 4-6, the same as that of the entire Ethereum transaction network. However, by calculating the average \textit{clustering coefficient} of the money laundering network corresponding to each case, we find that the money laundering network (both $\textsf{HeistEthNet}$ and $\textsf{HeistTokenNet}$) has a smaller clustering coefficient than the entire network. This may be because the special purpose of money laundering makes it lose the multi-hub and social characteristics of the entire network. Therefore, although the average shortest path length of the money laundering network is small enough, its clustering coefficient is small, so the money laundering networks do not exhibit the small world phenomenon.
% ys comment 上面一段最后一句的 we find that 删掉之后感觉前后语句会更顺【已修改~】
% 最后总结本小节的发现

% 合并到一起
% \noindent $\bullet$ \textbf{Finding 2.} In general, the self-loop ratio of crypto money laundering networks is lower than that of the entire network and the reciprocity is the opposite. The crypto money laundering network in Web3 is a frequent and dense sub-network, but does not exhibit the small world phenomenon.

% 大量基于智能合约的去中心化交易所提供的资产流动性和交易匿名性给反洗钱的检测带来了挑战。

% \begin{figure}%[bthp]
%     \centering
%     \includegraphics[width=0.9\linewidth]{Fig_motif_type_0930.pdf}
%     \vskip -2ex
%     \caption{Illustration of directed motifs: $M_1$ and $M_2$ are all connected two-node motifs; $M_3$ -- $M_{15}$ are all 13 connected three-node motifs; $M_{16}$ is the four-node bi-fan motif.}
%     % 
%     \vskip -2ex
%     \label{fig:motif_type}
% \end{figure}

% \begin{figure}[htbp]
%     \centering
%     \vskip -3ex
%     \includegraphics[width=0.7\linewidth]{Fig_motif_1001.pdf}
%     \vskip -2ex
%     \caption{Distribution of the various types of motifs in the money laundering network for different cases. The color of each motif in each case indicates the fraction of the count of motif in the case on a a linear scale (darker blue means a higher count).}
%     % 不同案件的洗钱网络中的各类模体的占比
%     % The color for the count of motif Mi,j indicates the fraction over all Mi,j on a linear scale—darker blue means a higher count
%     \vskip -2ex
%     \label{fig:motif_count}
% \end{figure}

\begin{figure}
    \centering
    %\vspace{-3ex}
    
    \subfigure[]{
			%\centering
			\includegraphics[width=0.2\linewidth]{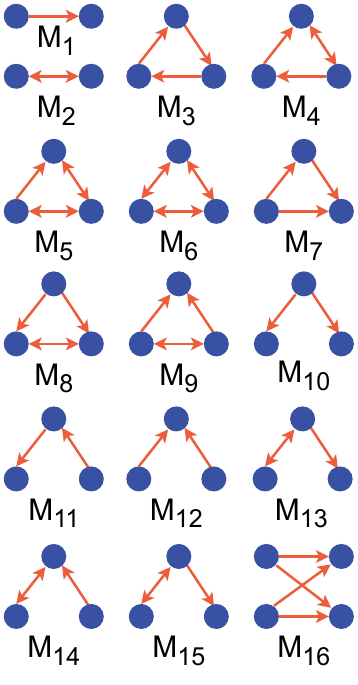}
	}
	%\hspace{-1ex}
	%\vspace{-1.5ex}
	\subfigure[]{
        \includegraphics[width=0.5\linewidth]{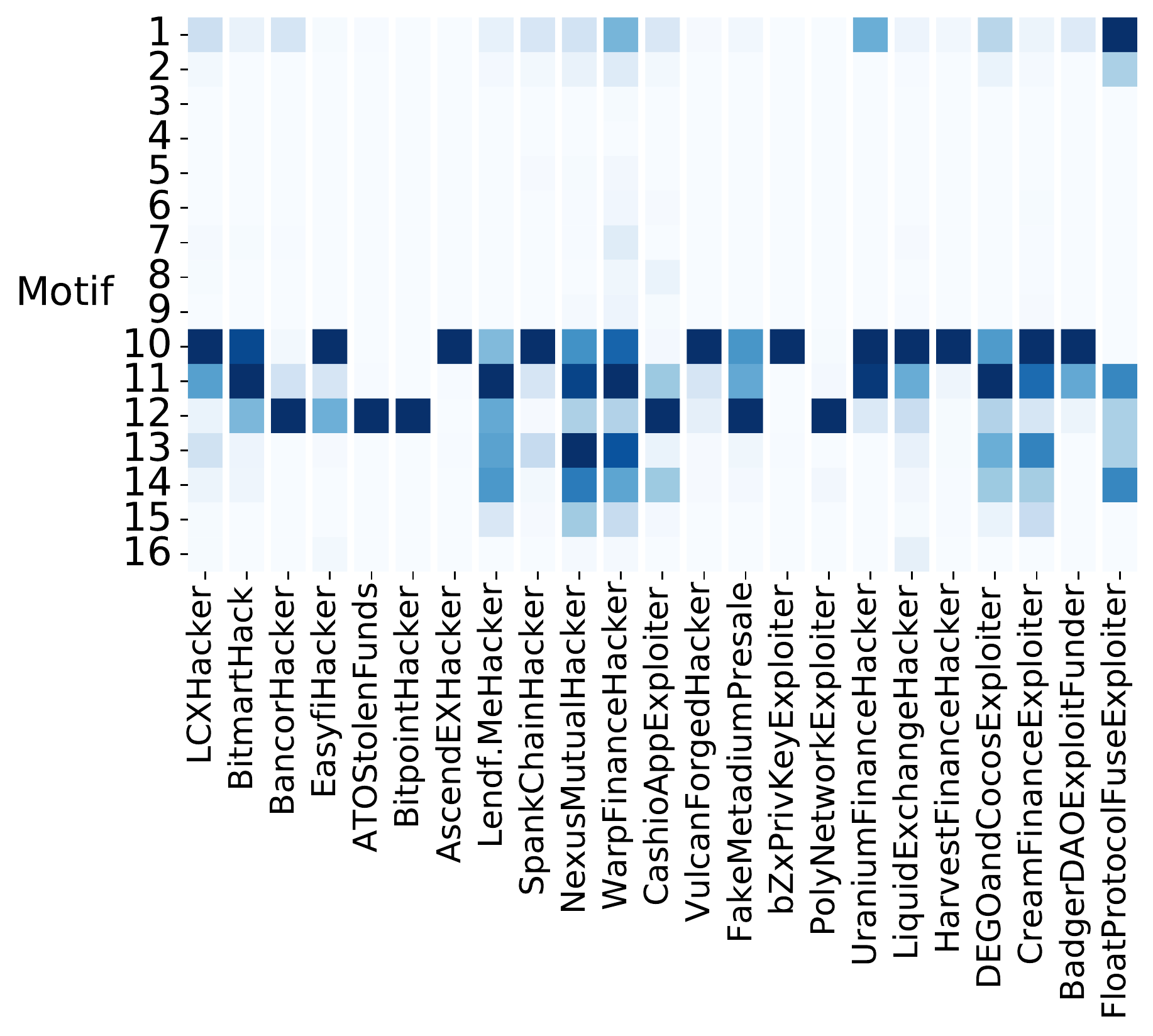}
    }
    %\vspace{-1.5ex}
    \caption{(a) Directed motifs: $M_1$ and $M_2$ are all connected two-node motifs; $M_3$ -- $M_{15}$ are all 13 connected three-node motifs; $M_{16}$ is the four-node bi-fan motif. (b) Distribution of the various motifs in money laundering networks.}
    %The color of each motif in each case indicates the fraction of the count of motif in the case on a a linear scale (darker blue means a higher count).

    \label{fig:motif_both}
    
    %\vskip -2ex
    
\end{figure}

\subsection{High-order Motifs Counting}
% 高阶结构是由网络主题~cite{Benson2016Motif}捕获的，它是网络中反复出现的小子图。为了描述高阶互动模式，我们统计了每个案例中洗钱网络中2节点、3节点和双扇形连接的有向图案（如图~ref{fig:motif_type}所述）的百分比。图~ref{fig:motif_count}显示了23个案例中每个motif的百分比结果（其他的遇到了内存外的错误）。我们可以看到，有些图案明显占优势。
Higher-order structure of networks can be captured by network motifs~\cite{Benson2016Motif} which are recurring small subgraphs in the network. To characterize higher-order patterns, we count the percentage of directed motifs (described in Figure~\ref{fig:motif_both}(a)) of the simple, directed money laundering network of each case. Figure~\ref{fig:motif_both}(b) shows the results for the percentage of each motif in 23 cases (the others encountered out-of-memory errors). 
%We can see that some motifs are clearly dominant.
% 然后，我们将实验结果与整个以太坊区块链网络~cite{Lee2020Measurements}的图案统计进行比较，得到一些有趣的观察结果。
Then, we compare with the entire Ethereum blockchain network~\cite{Lee2020Measurements} and obtain some interesting observations: %It is reported that the motifs with largest density observed are primarily chain and closed triangles in the entire network ~\cite{Lee2020Measurements}. However, 
\begin{enumerate}[(i)]

\item The fractions of \textit{closed triangular motifs} are quite low ($M_3$--$M_9$) in money laundering networks. This may be because the pattern of closed triangle motifs is a manifestation of assets circulating internally, such as wash trading behavior~\cite{victor2021detecting}, which is not consistent with the intent to launder assets.
% (i) 据报道，在整个网络中观察到的密度最大的图案主要是链和封闭三角形~cite{Lee2020Measurements}。然而，在洗钱网络中存在的封闭三角形图案的比例相当低（M_3$-M_9$）。这可能是因为封闭三角形图案是资产在内部流通的表现，如洗牌行为~cite{victor2021detecting}，这与转移资产的洗钱意图不一致。

\item On the contrary, \textit{open triangle motifs} are the most frequent motifs that appear in the money laundering network, of which there are three most, i.e. $M_{10}$-$M_{12}$. These three motifs correspond exactly to three phases of money laundering: $M_{10}$ belongs to the placement phase, which spreads the illegally obtained stolen money and extends the money path; $M_{11}$ belongs to the layering phase, which continuously passes stolen funds and makes it more difficult to trace; $M_{12}$ belongs to the aggregation phase, which collects the scattered laundered stolen money for withdrawal.

% 这三种模体恰好对应洗钱的三个阶段。M10属于放置阶段，散播非法获得的赃款；M11属于分层阶段，不断传递赃款，增加追踪的难度；M12属于聚合阶段，汇总分散洗钱的赃款进行提现。
\item In particular, the money laundering network of DeFi exploit cases has more $M_{13}$-$M_{15}$ motifs, in which the bidirectional edges are most likely related to a classic DeFi action -- token trade (also called exchange), i.e., the trader's account sells a certain amount of a certain token in exchange for a certain amount of another token in a liquidity pool of an Automated Market Maker (AMM). 
To this end, we identify the DeFi token swap action in each case, referring to \textsc{DeFiRanger}~\cite{Wu2021DeFiRanger}. We find that the number of $M_{13}$-$M_{15}$ motifs does have a strong correlation with the DeFi token swap action. For example, at least 70 token swaps were identified in money laundering of Cream Finance Exploiter~\footnote{\url{https://etherscan.io/address/0x24354d31bc9d90f62fe5f2454709c32049cf866b}}, and its $M_{13}$-$M_{15}$ motif fractions are also relatively high.
% (3) 特别地，与DeFi exploit案件洗钱网络有更多的M13~M15模体，这很可能是和一种经典的DeFi行为有关--代币兑换，即，交易者的账户卖出一定数量的某种代币，换取一定数量的另一种代币。
% 为此，我们参考吴的方法[]，识别并统计了每个案件中出现DeFi 代币兑换的次数。例如，在Harvest Finance Exploiter的洗钱网络中发现了221个代币互换，在Cream Finance Exploiter中也发现了70个代币互换。
\end{enumerate}

%\subsection{\red{Cross-asset analysis}}
% Add by Qishuang.
% To Qishuang.
% （1）以避免资产冻结: 如USDT 的发行者可能会冻结非法钱包持有的资产。因此，犯罪分子使用DEXs将可冻结的资产交换为不可冻结的资产。例如，2021年12月发生的AscendEX Exploit，攻击者通过Curve.Fi服务将盗取的价值5.7 million美元的USDT，花了大约2小时40分钟迅速换成了代币DAI USDC。但，也有的黑客没有及时将被盗资产兑换为不可冻结资产，
%如EasyFi Exploit在十小时后才将盗取的USDT在1inch DEX聚合器兑换。【注：这里提到的两个案例来自于Elliptic报告原文的P14-15，恰好这两个案例也在咱们本文中有数据，比较合适。】
% （2）将代币交换为ETH: 通常是为通过Tornado Cash发送做准备。BitmartHack案件中，黑客通过1 inch去中心化交易所将MANA代币换为了ETH，然后，send swapped ETH to Tornado.Cash for mixing【注：Elliptic报告中提的案例在本文数据集没有出现，所以需要齐双在咱们数据集里先找找。Elliptic报告原文的P16】
% （3）兑换代币以准备桥接到其他区块链。例如，NexusMutualHacker案例的被盗资产被黑客兑换为renBTC -以太坊上的wrapped版本的BTC，然后可以使用RenBridge 桥接到比特币区块链。
%或者某某案例的被盗资产被黑客兑换为anyXXX - 以太坊上的XXX链代币的形式，然后可以使用AnySwap进行跨链到XXX公链上。【注：Elliptic报告中提的案例在本文数据集没有出现，所以需要齐双在咱们数据集里先找找。Elliptic报告原文的P17】

%我们更进一步地探索黑客洗钱的Illicit Token Flows。我们分析了本文这33个案例，发现犯罪分子窃取了923类不同的代币资产。这些不同类型的资产经过多种DEXs的token swap之后(在某些情况下发生多次)，较为热门的目的地token是：ETH、USDT、ETH和DAI。发生跨资产行为的平均时间是开始清洗被盗资产后的15小时。较为热门的DEXs服务包括Uniswap、1 inch等。【注：以上结论来自Elliptic报告，需要在我们的数据集上探究，是否会偏差很大。Elliptic报告原文的P18】

\red{To further understand the criminal activities of hackers using DEXs/AMMs, we explore and analyze the cross-asset behavior of hackers in the money laundering process and its purpose, and attribute it to the following activitie:}
% 为了更进一步地了解黑客利用DEXs/AMMs进行的犯罪活动，我们探索和分析了黑客洗钱过程中的跨资产行为及其目的，并归结为了以下几种活动
\begin{enumerate}[(i)]
    \item \red{\textbf{\textit{Swapping tokens to non-freezable assets.}} For example, Tether (USDT) is a stablecoin pegged to the US Dollar, operated by Tether Limited Inc. USDT issuers may freeze assets held by illegal addresses. As a result, criminals use DEXs to swap freezable assets for non-freezable ones. For example, in the AscendEX Exploit\footnote{\url{https://etherscan.io/address/0x2c6900b24221de2b4a45c8c89482fff96ffb7e55}} event that occurred in December 2021, the attackers quickly exchanged \$5.7million worth of USDT stolen through the Curve.Fi service for DAI, USDC, in about two hours and 40 minutes.}
    % （1）以避免资产冻结: 如USDT 的发行者可能会冻结非法钱包持有的资产。因此，犯罪分子使用DEXs将可冻结的资产交换为不可冻结的资产。例如，2021年12月发生的AscendEX Exploit，攻击者通过Curve.Fi服务将盗取的价值5.7 million美元的USDT，花了大约2小时40分钟迅速换成了代币DAI USDC。但，也有的黑客没有及时将被盗资产兑换为不可冻结资产 % to avoid asset freezing

    \item  \red{\textbf{\textit{Swapping tokens for mixing.}} Many criminals make use of DEXs to swap their stolen tokens to ETH for mixing. For example, in the Bitmart Hack\footnote{\url{https://etherscan.io/address/0x39fb0dcd13945b835d47410ae0de7181d3edf270}} event that occurred in December 2021, the criminal swapped MANA token for ETH in 1 inch DEX, then sent swapped ETH to Tornado.Cash for mixing.}
    % （2）将代币交换为ETH: 通常是为通过Tornado Cash发送做准备。BitmartHack案件中，黑客通过1 inch去中心化交易所将MANA代币换为了ETH，然后，send swapped ETH to Tornado.Cash for mixing
    
    \item  \red{\textbf{\textit{Swapping tokens to bridge them to other blockchains.}} Cross-chain transactions of criminals are cunning behavior to confuse the flow of dirty money. Before Cross-chain transactions, criminals need to swap assets for tokens convertible on bridges. For example, in the Nexus Mutual Hacker event that occurred in December 2020, the stolen ETH was swapped for renBTC, then the renBTC was bridged to the Bitcoin blockchain. RenBTC is a wrapped version of bitcoin on Ethereum which can then be bridged across to the Bitcoin blockchain using RenBridge. }
    % （3）兑换代币以准备桥接到其他区块链。例如，NexusMutualHacker案例的被盗资产被黑客兑换为renBTC -以太坊上的wrapped版本的BTC，然后可以使用RenBridge 桥接到比特币区块链。

\end{enumerate}

\red{We go a step further to explore Illicit Token Flows for money laundering. We analyze these 33 cases in this paper and find that criminals stole 923 different types of token assets. These different types of assets went through multiple DEXs token swaps (in some cases occurring multiple times), with the more popular destination tokens being: ETH, USDT, WETH, and DAI. the average time for cross-asset behavior to occur was 15 hours after the start of laundering the stolen assets. Some of the more popular DEXs services include Uniswap, 1 inch, etc.
}

\begin{figure*}[t]
    \centering
    %\vspace{-2ex}
    \subfigure[Word cloud of service providers]{
			%\centering
			\includegraphics[width=0.4\linewidth]{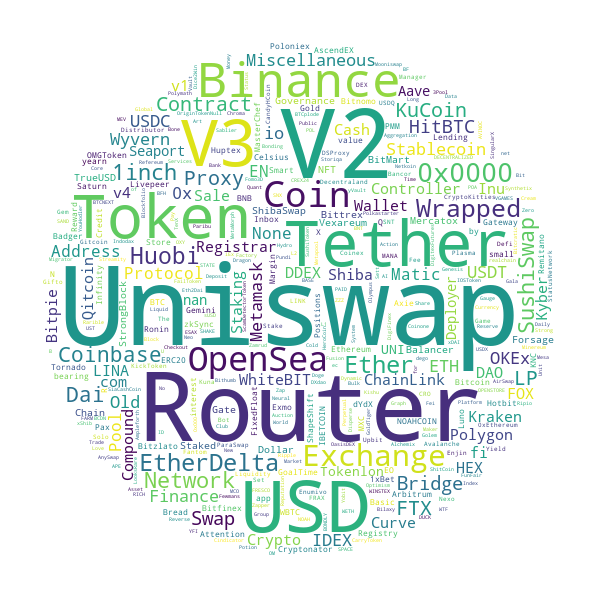}
	}
	%\hspace{-1ex}
	%\vspace{-1ex}
	\subfigure[Evolution of destination service providers]{
        \includegraphics[width=0.55\linewidth]{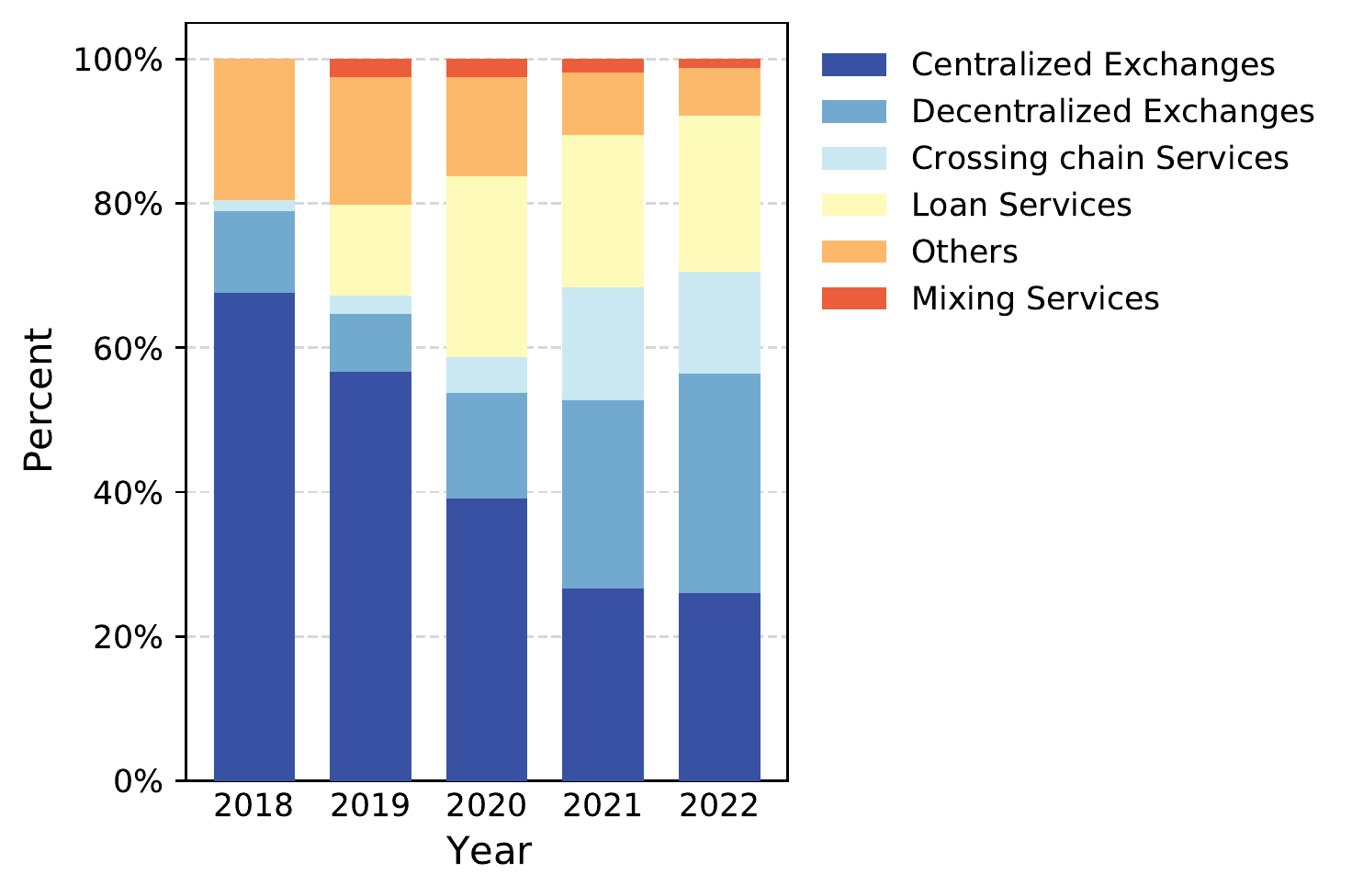}  
    } %服务商种类堆叠柱状图_1013.pdf
    %\hspace{-1ex}
    %\vspace{-1ex}
   
    \caption{Illustration of the economic impact of money laundering in terms of destination service providers and marketplace.}
    %\caption{(a) Word Cloud for descriptive texts of the Service Providers. (b) Destination of stolen funds from 2018 to 2022. (c) Tx volume from KucoinHacker vs. ETH price listed on Etherscan }
    \label{fig:cloud_percentage_price}
    %\vspace{-2ex}
\end{figure*}

\vspace{1.5ex}
\begin{mdframed}
    \textbf{Finding 2.} In general, the self-loop ratio of crypto money laundering networks is lower than that of the entire network and the reciprocity is the opposite. The crypto money laundering network in Web3 is a frequent and dense subnetwork, but does not exhibit the small world phenomenon. There exists a large number open triangle interaction patterns but few closed triangle patterns in crypto money laundering networks. Particularly, the open triangles in DeFi exploit cases contain more bidirectional edges, reflecting the method of further obfuscating stolen assets through Defi's token exchange. 
    The use of DEXs/AMMs by crypto criminals is closely associated with exploits in the DeFi projects and hacks of exchanges. 
\end{mdframed}
\vspace{1.5ex}

% % 最后总结本小节的发现
% % 黑客如何在洗钱中混淆来源？洗钱账户群体的交互特征？
% \noindent $\bullet$ \textbf{Finding 3.} There exists a large number open triangle interaction patterns but few closed triangle patterns in crypto money laundering networks. Particularly, the open triangles in DeFi exploit cases contain more bidirectional edges, reflecting the method of further obfuscating stolen assets through Defi's token exchange. 

    % 黑客控制的洗钱账户群体中几乎没有闭三角的交互模式，而更多地是通过开三角的交互模式进行更快速洗钱，以达到分散资金、延长链路、聚集提现的目的。其中，DeFi被攻击项目的开三角中含有更多的双向边，反映了黑客通过Defi的代币兑换的方法来进一步混淆被盗资产。

\section{RQ3: Economic Impact of Laundering}

\label{sec:economic}

% zj comment: 感觉evolution of Service Providers并不是Economic Impact of Laundering，内容和节标题不太对应，要不把这一小节放7.3弱化一下存在感.【7.1和7.2是有关系的，7.1发现脏钱的去向之一是各种交易平台，那么7.2继续探究这些流入交易平台的脏钱是否影响币价，这个递进关系现在确实没写出来，我在7.2的开始补充这个承上启下关系了】
\subsection{Service Providers of Laundering Exit}
%Writer: Qishuang Fu
%安全事件发生之后，几乎所有的黑钱都流入了服务商，被洗.所以研究各种服务商的占比和变化是很重要的。

%介绍词云：
    %参考伟利师兄对词云的描述：
    %为了产生第一印象，我们画了一个收集的案件的服务商的词云图。
    %可以看到，最频繁出现的字是Uniswap。
    %介绍Uniswap:很有名的点对点的去中心化交易所，使得用户不需要中心化第三方就可以交易。所以，他提供了宽广平台给犯罪分子洗钱。t=
    %此外，还存在一些其他的服务商，比如binance , opensea , sushiswap.
%介绍服务商随时间的变化：
    %为了进一步探究参与洗钱的服务商随着时间的变化，我们将服务商分成了六类，分别是，然后画了一个图of destination of stolen founds from 2018 to 2022。
    %正如图中所示，各种服务商的占比随着时间在变化的。
    %中心化交易所的占比在2018年是巅峰，但是从2019年开始就在下降，这是因为中心化交易所加强了自己的AMK和KYC的业务。
    %去中心化交易所的占比在增加，从中可以推测去中心化交易所更有可能逃离法律监管。
    %跨链服务占比在逐年增长从2019年，这说明犯罪分子越来越狡猾，在多条链进行洗钱
    %%借贷服务是一种相对被动的投资方式。用户将他们的钱放入借贷服务中心，他们会受到投资的汇报，可能只有原始资产的一小部分。对于犯罪分子来说，借贷服务不仅提供了几千的机会，还帮助他们赚取额外的收益。
    %此外，还有一些钱是通过mixing services like tornado.cash去混淆来源的。
    %还有一些其他的服务商,比如犯罪分子将加密货币存入Air Wallet，这是一种分布式空投和数字钱包平台。  
In the aftermath of Web3 security incidents, almost all black money flows to service providers to be washed. Thus, it is necessary to present the percentage and changes of various service providers.
For a first impression, in Figure~\ref{fig:cloud_percentage_price}(a), we draw a word cloud graph of the
service providers involved in collected events. As we can see, the most frequent word is ``Uniswap''. Uniswap is a decentralized exchange of great name that enables peer-to-peer market making and enables users to trade or swap cryptocurrencies without any involvement with a centralized third party, so it provides a wide platform for criminals to money laundering. Additionally, there are some other typical service providers popular among crypto laundering in Web3. For example, Binance (an eminent centralized exchange), Opensea (the largest NFT marketplace), and SushiSwap (a decentralized exchange similar to Uniswap).

To further explore the evolution of service providers involved in money laundering over time, we first divide the service providers into six categories, which are centralized exchanges (CEXes), decentralized exchanges (DEXes), crossing chain services, loan services, mixing services, and others.
Then, we draw a stacked bar chart displaying the percentage of various service providers change over time as shown in Figure~\ref{fig:cloud_percentage_price}(b). 
% As shown in Figure~\ref{fig:cloud_percentage_price}(b), the percentage of various service providers changes over time. 
%1008by蕴湄：To further explore the evolution of service providers involved in money laundering over time, we divide the service providers into six categories which are centralized exchanges (CEX), decentralized exchanges (DEX), crossing chain services, loan services, mixing services, and others, then we draw a figure displaying the percentage of various service providers change over time as shown in Figure~\ref{fig:cloud_percentage_price}(b). 【get】

As can been seen, the preferences of the service providers to which this dirty money goes change over time.
On the one hand, centralized exchanges, once the top destination for stolen funds in 2018, phased down in 2019. The reason may be that CEXes have enhanced AML and KYC procedures at the request of the regulatory section in recent years~\cite{fatf2019VASP_old,fatf2021VASP}. On the other hand, there is an increase in the share of DEXes, which can infer that DEXes without a centralized third party is more likely to escape law enforcement investigations. Moreover, the share of crossing chain services is growing year by year since 2019, which allows black money to circulate and confuse on multiple chains, indicating that criminals are becoming more crafty.
There is even dirty money flowing to lending services such as Aave, Compound, Dydx, etc. By using the liquidity pool of lending services, criminals can not only conceal the source of dirty money and reduce the possibility of being traced, but also earn extra income via providing huge amount dirty money to liquidity pools. 
%This is really killing two birds with one stone for criminals.
% ys comment 上面一段只有on the other hand没有on the one hand【已修改，在前面一句加上了On the one hand】
% 甚至还有脏钱流向Aave、Compound、Dydx等借贷服务。通过利用借贷服务的流动性池，犯罪分子不仅可以掩盖脏钱的来源，减少被追踪的可能性，还可以通过向流动性池提供大量脏钱来赚取额外收入。这对犯罪分子来说确实是一石二鸟。
%\textcolor{red}{Loan services devote to relatively passive forms of investment, such as Aave, Compound, Dydx, etc. To be specific, users send their cryptocurrencies to a loan service or staking in liquidity pools, in return they will receive a consistent return which may be just a small percentage of their origin assets. So for criminals, loan services not only offer opportunities for money laundering but also help them earn extra profits.}
%Besides, portion of all funds stolen are passed through mixing services like Tornado.Cash to obscure their illicit origins. 
%自从2019年混币服务Tornado Cash成立，他就一直成为脏钱的去向。这能推测混币服务是一个经典且效果好的洗钱服务；难怪Tornado Cash美国财政部制裁。
In addition, since the inception of the mixing service, Tornado.Cash, in 2019, it has been one of the destinations of dirty money. It can be presumed that it is a classic and effective money laundering service, no wonder Tornado.Cash was recently sanctioned by the U.S. OFAC\footnote{\url{https://home.treasury.gov/news/press-releases/jy0916}}.
In addition, there exist some other kinds of service providers. For example, criminals deposit crypto to Air Wallet, a kind of distributed airdrop and digital wallet platform.

% \begin{figure}%[htbp]
%   \centering
%   \includegraphics[width=0.5\linewidth]{Fig_new_service_cloud.png}
%   \vskip -1.5ex
%   \caption{Word Cloud for descriptive texts of the Service Providers.}
%   \vskip -1.5ex 
%   \label{fig:cloud}
% \end{figure}

% \begin{figure}%[htbp]
%   \centering
%   \includegraphics[width=0.7\linewidth]{Fig_services_year_1005.png}
%   \vskip -1.5ex
%   \caption{Destination of stolen funds from 2018 to 2022.}
%   \vskip -1.5ex 
%   \label{fig:percentage}
% \end{figure}

%\vspace{15em}

% 最后总结本小节的发现
%被用来洗钱的服务商随着时间变化。去中心化交易所、跨链服务和借贷服务的份额逐年增加的现象说明犯罪分子反监管的意识越来越强，在不断地寻求更加狡猾和更加隐秘的洗钱手段。
% \noindent $\bullet$ \textbf{Finding 4.} The service providers being used to launder money are changing over time. The phenomenon of decentralized exchanges, cross-chain services, and lending services increasing their share year by year indicates that criminals are becoming more anti-regulatory and are constantly seeking more cunning and stealthier means of money laundering.

% This not only increases the risk of DeFi project assets being frozen, but also has a negative impact on the reputation and development of the Web3 ecosystem. 
\begin{figure}
  \centering
  %\vskip -1.5ex
  \includegraphics[width=0.7\linewidth]{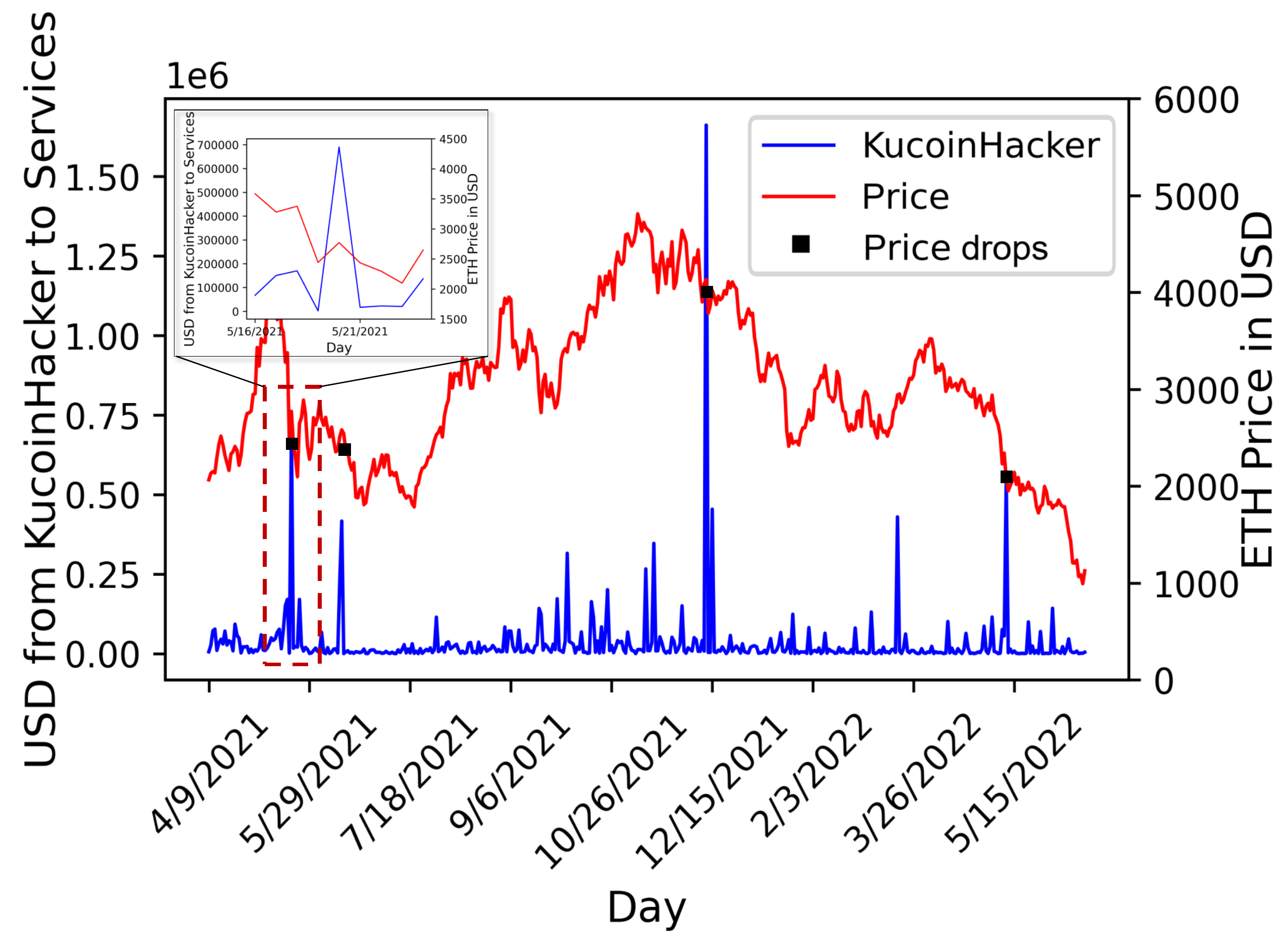}
  %\vskip -2ex
  \caption{Transaction volume from Kucoin Hacker vs. ETH price}
  %\vskip -1.5ex 
  \label{fig:price_volume}
\end{figure}
%\newpage
%\vspace{-1.5ex}
\subsection{Crypto Price Drop Caused by Cash-outs} 
% \subsection{Price fluctuation} 

% \begin{figure}%[htbp]
%   \centering
%   \includegraphics[width=0.7\linewidth]{ETH_Amount_Price_KucoinHacker_2plot.png}
%   \vskip -1.5ex
%   \caption{Tx volume from KucoinHacker vs. ETH price listed on Etherscan}
%   \vskip -1.5ex 
%   \label{fig:percentage}
% \end{figure}

%洗钱账户在大量抛售Stolen crypto提现时，极大可能会影响币价的波动。因此，我们探究黑客将以太币send给各种服务商是否会影响ETH的币价。

%由于空间限制，我们只展示其中一个典型案件-Kucoin的结果.
%从图中可以看出，黑客将以太币send给各种服务商确实会影响ETH的币价。

%具体来说，我们可以看到KucoinHackers 发送ETH to service providers from April 2021 to May 202Ma2, 并且存在5个显著的交易量峰值， for example, May 2021, December 2021, March 2022 and May 2022. 当交易峰值出现时，说明hack在大量的提现，ETH的价格在显著下降。因此，我们可以推测，cashouts cause increased volatility in Bitcoin’s price, and that they correlate significantly with Bitcoin price drops，这可能是因为黑客急于提现，低价售卖ETH，导致ETH价格降低。
In the previous section, we find that various trading platforms are destinations for dirty money.
There is a great possibility that money laundering accounts selling stolen crypto in large quantities to withdraw cash will affect the volatility of the crypto price. 
Therefore, in this part, we further explore whether hackers sending ETH to various service providers affects the price of ETH.
Due to space limitation, here we only show the result of one of the typical cases - Kucoin Hacker\footnote{\url{https://etherscan.io/address/0xeb31973e0febf3e3d7058234a5ebbae1ab4b8c23}}. From Figure~\ref{fig:price_volume}, we see that KucoinHackers sent ETH to service providers from April 2021 to May 2022 and there exist 5 apparent spikes of transaction volume, for example,  May 2021, December 2021, March 2022 and May 2022. When a transaction spike occurs, which means that the hacker is withdrawing a lot, the price of ETH drops significantly. Therefore, we can presume that a large number of cashouts correlate significantly with the price of ETH drops. This may be because hackers are eager to withdraw cash and then sell ETH at low prices, resulting in a significant drop in the price of ETH.
%we see that PlusToken wallets sent a steady flow of Bitcoin starting in mid-April and spiking just before the arrests in late June
%cashouts cause increased volatility in Bitcoin’s price, and that they correlate significantly with Bitcoin price drops

% 另外，我们发现NFT被盗后也面临被低价抛售的命运。Arthur热钱包被盗的案发当日，黑客就直接将盗取的17个Azuki系列NFT，以10ETH左右的价格出售或拍卖，明显低于当时平均市场价格（13ETH）。其中价格降幅最大的是Azuki #606——从被盗前的78 ETH ($105,506.70)降到50.15 WETH ($67,835.40)被抛售。尔后，黑客为了防止被冻结，先将被盗的NFT转移到其他钱包再进行抛售。这时候，我们更加难区分这个购买NFT的对象是黑客控制的账户还是普通用户，增大了追踪黑客洗钱交易的难度。另一方面，黑客以低价对被盗NFT进行抛售，可能会导致NFT项目的信任危机。对于普通用户来说，很可能不小心购买了这些被盗NFT，导致其后续被交易平台列为黑名单，不利于NFT市场稳定。

%OpenSea在窃盗案发生很久后，才对某些资产进行标记，这对买家很不公平。今年7月初，就有一名OpenSea发推抱怨，他在买下CloneX NFT的88天后，该NFT却被OpenSea标记为可疑。

In addition, we find that the stolen NFTs also face the fate of being sold at low prices. On the day of the Arthur Hot Wallet heist, the hacker directly sold or auctioned off the 17 stolen Azuki NFTs for around 10 ETH, which was significantly lower than the average market price (13 ETH) at the time. One of the biggest price drops was Azuki\#606\footnote{\url{https://etherscan.io/nft/0xed5af388653567af2f388e6224dc7c4b3241c544/606}} - from 78 ETH before the heist to 50.15 ETH when it was sold off. Moreover, the hacker transferred the stolen NFT to other wallets before dumping it in order to prevent it from being frozen. On the one hand, it is not obvious to distinguish whether the purchaser of the stolen NFT is a hacker-controlled account or an ordinary user, increasing the difficulty of tracing the hacker's money laundering transactions.
On the other hand, ordinary users are likely to accidentally purchase these stolen NFTs, resulting in their subsequent blacklisting by the trading platform, which is not conducive to NFT market stability.
% ys comment 上面一段it is more difficult ... 这一句和making it more difficult句式重复，可以换一个表达【有道理，已修改】
% ys comment 同样只有on the other hand，没有 on the one hand【已修改，把上一句原来的“At this point”改为“On the one hand”】
% zj comment: 上面一段几个实例需要加一些引用或者图【get，加footnote。】

% making it more difficult to trace the money laundering transactions. 
% Azuki\#606

% \subsection{被低价抛售的NFT？} 

% 最后总结本小节的发现
\vspace{1.5ex}
\begin{mdframed}
\textbf{Finding 3.} 
A large number of cashouts of hackers correlate significantly with ETH price drops. The hackers send lots of stolen ETH to service providers in order to get back clean fiat currency. Hackers dumping stolen NFTs at low prices could lead to a crisis of trust in the NFT programs and increases the risk of freezing the crypto assets of a genuine user of NFT trading platforms.
The service providers being used to launder money are changing over time. The phenomenon of decentralized exchanges, cross-chain services, and lending services increasing their share year by year indicates that criminals are becoming more anti-regulatory and are constantly seeking more cunning and stealthier means of money laundering.
\end{mdframed}
\vspace{1.5ex}

% \noindent $\bullet$ \textbf{Finding 5.} A large number of cashouts of hackers correlate significantly with ETH price drops. The hackers send lots of stolen ETH to service providers in order to get back clean fiat currency. %Usually, they may set a relative low price to sell the crypto quicker, so it will make the price of ETH lower. 
% Hackers dumping stolen NFTs at low prices could lead to a crisis of trust in the NFT programs and increases the risk of freezing the crypto assets of a genuine user of NFT trading platforms.
% % NFT被盗后被抛售，可能导致NFT市场不稳定，增加普通用户购买NFT资产被冻结的风险。

% \newpage
%\vspace{-1.5ex}
\section{Counterfeit Token Deployment for Cunning Laundering}

\begin{figure}%[htbp]
  \centering
  \includegraphics[width=0.8\linewidth]{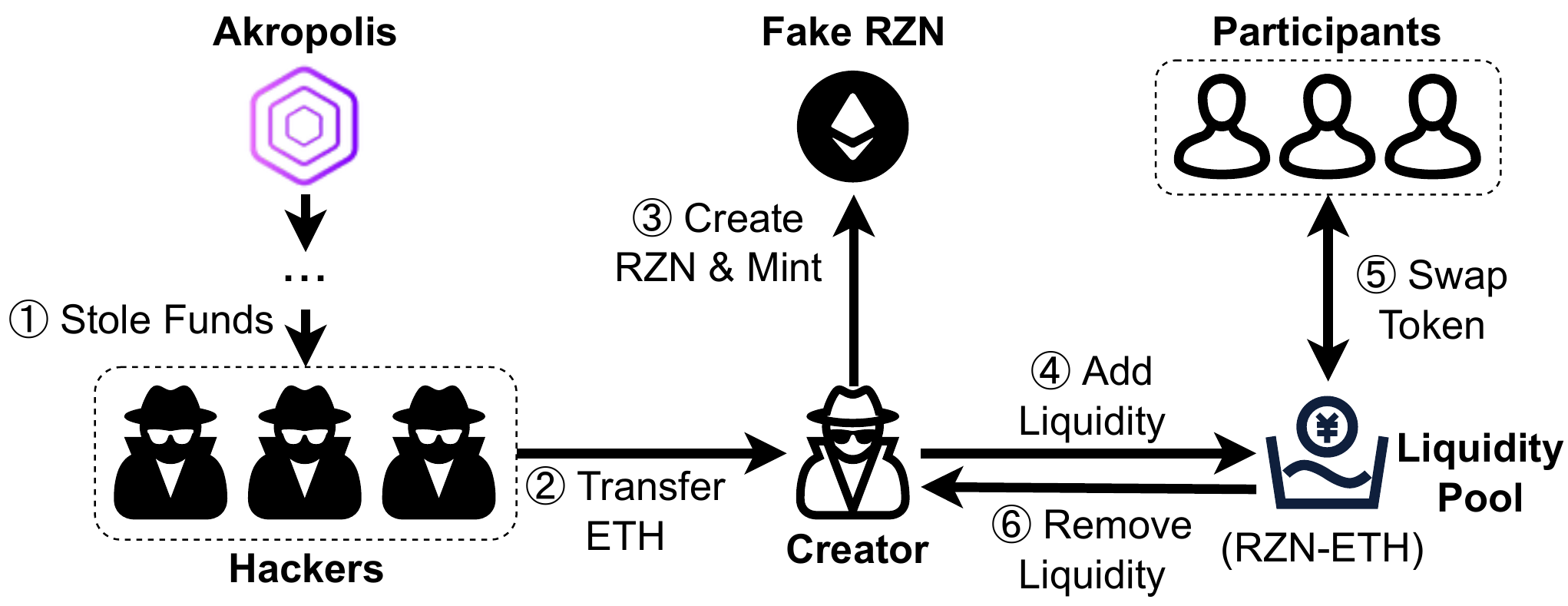}
  %\vspace{-2ex}
  \caption{The Role Fake Tokens Play in Money Laundering.}
  
  %\vspace{-2ex}
  
  \label{fig:faketoken}
\end{figure}

In addition to money laundering techniques such as layered transfers and cross-chaining, more cunning hackers may disguise themselves as other common players to evade detection in the Web3 ecosystem, where the existence of counterfeit tokens provides the perfect opportunity for hackers to launder money. Researchers~\cite{gao2020tracking,xia2021trade} have found that counterfeit tokens are prevalent in the Web3 ecosystem because most DEXes do not enforce any rules for token listing. Hackers can easily create counterfeit tokens and liquidity pools, and even disguise themselves as ordinary speculators to launder illicit funds from liquidity pools of counterfeit tokens.
To this end, we conduct an empirical analysis based on the counterfeit token dataset provided by Gao \textit{et. al.}~\cite{gao2020tracking}, and are surprised to find that in \textbf{13 of the 33} cases in this paper are related to counterfeit tokens some way.
One notable case is Akropolis Hacker\footnote{\url{https://etherscan.io/address/0x9f26ae5cd245bfeeb5926d61497550f79d9c6c1c}} (a DeFi Hacker). By tracing downstream laundering transactions of Akropolis Hacker, we find evidence that this hacker was laundering money by creating counterfeit tokens. As plotted in Figure~\ref{fig:faketoken}, the main procedures are as follows:
\begin{enumerate}[(i)]
    \item Using the tracing method mentioned earlier, we see that the hacker cascaded the stolen funds from Akropolis to several accounts under his control (0x1c80\footnote{\url{https://etherscan.io/address/0x1c80f8670f5c59aab8e81e954aabb64dabde2710}} and 0x982d\footnote{\url{https://etherscan.io/address/0x982dd33d6bc5 bf83eedcbcab92e4899c7a}} identified as ``Heist'' in our dataset), and transferred funds to another controlled account 0x1f8\footnote{\url{https://etherscan.io/address/0x1f84ba7bacd29e875367688b38ecccb7849b50fa}}.
    \item Subsequently, address 0x1f8 created the \textbf{fake token RZN} (Rizen Token)\footnote{\url{https://etherscan.io/token/0x9c91310c9bf1c779b667f46322d33bfdc96c1a07}}, and a liquidity pool on Uniswap\footnote{\url{https://etherscan.io/address/0x658b4a15aae288757c41a9b074ab1881d3ecad0c}} with 594 fake RZN and 0.877 ETH.
    \item The hackers then manipulated the liquidity pool through multiple accounts, posing as ordinary speculators and participating in the trading of counterfeit tokens, e.g. address 0x8c4d\footnote{\url{https://etherscan.io/address/0x8c4dedecbe3e8fbcc0501599cb59e7feadd99ffc}} sold 1000 fake RZN and got 300 clean ETH. 
    \item Finally, the RZN creator removed the liquidity of 2144 RZN coins and 0.25 ETH. The hackers successfully laundered the illegal funds by disguising their addresses as ignorant participants.
\end{enumerate}

% 黑客通过伪装地址成无知投机者的身份，将非法资金成功洗白。

%%%%%%%%%%%%%%%%%%%%%%%%%%%%%%%%%%%%
% 2023-01-06----------
%\red{\textbf{[Bytecode-based Suspicious Contract Detection. Add by Kailin.]}}
% 上述分析过程只能识别出已被标记的scam token，然而，在洗钱过程中还有许多其他的假冒代币来进行伪装。从监管的角度来说，识别这些假冒的代币或者DeFi项目是很重要的。因此，在下面，我们试图基于合约相似度检测方法，来识别更多被黑客用于洗钱的可疑假冒token和DeFi。收到论文Analyzing Ethereum’s Contract Topology的启发，我们基于合约相似度的检测方法如下：

% （1）我们找到交易发起方标记为Heist的交易，根据该交易contractAddress字段的地址，从Xblock公开数据中获取其合约字节码，作为待检测合约数据集。同时，上文提到了论文Trade Or Trick的合约字节码作为已知假冒token合约数据集。
% （2）我们反编译所有反汇编所有合约，得到opcode并去掉所有参数,但保持操作码出现的顺序；计算n-gram（用n=5），使得每个合约有一个高维的向量表示
% （3）我们将待检测合约数据集与已知假冒token合约数据集中的每个数据都两两组成一对，计算合约的相似度，若合约相似度大于预先计算的阈值，则认为待检测合约为可疑假冒token合约。具体地，用两个合约的n-gram向量的余弦相似度计算（cosine similarity），得到一个介于0(完全不同)和1(完全相同)的相似度结果。在本文中，我们设定相似度阈值为0.90

% 通过实验我们发现，在本文的洗钱数据集中，还新发现了**个可疑的假冒token合约，属于案例a,b,c。我们人工验证了相似度最高的三个合约，……
\vspace{1.5ex}
\begin{mdframed}
\textbf{Finding 4.} 
Hackers launder money anonymously by creating fake tokens and disguising their addresses as ignorant participants, which is an upgraded method of money laundering in Web3. At the same time, the fake tokens created by hackers simultaneously increase the risk of ordinary users falling for the scams.
\end{mdframed}
\vspace{1.5ex}
\section{Money Laundering Core Group Identification. }
% Add by Yunmei.
% 这【注：这部分是基于Fraudar算法的应用，但是Fraudar本身不是反洗钱的算法而是异常检测的算法，所以讲法上需要往反洗钱上靠。】

% 本文在前面数据构成的章节介绍了如何根据洗钱的起点，挖掘下游未知的洗钱账户和网络。
In the previous section of dataset construction, we introduce how to mine downstream unknown money laundering accounts and networks based on the starting hackers, exploiters, and scammers. Here, the target of the identification of core money laundering organizations is to find core account groups with more intensive interactions and frequent capital transactions in the relatively sizeable downstream money laundering network, as a supplement to criminal evidence collection.

% 为了达到这个目标，在本文原始爬取到的洗钱网络的基础上，我们首先定义账户的洗钱可疑度，并利用近似贪心算法对原始下游网络进行剪枝，同时构建了最小优先树来加速迭代的过程，最终得到可疑度最高的子网络作为洗钱的核心团伙。

%可疑度的核心思想来源于RQ1和RQ2观察到的洗钱特征：洗钱者会创建了一个大量而密集的转账子图，这是因为洗钱账户为了避免被发觉和冻结，需要把大量的资金在短时间内转移到各类服务商和流出交易平台，最终导致了一个密集的转账子图。
In order to achieve this target, on the basis of the money laundering network initially crawled in this paper, we first define the money laundering \textit{suspiciousness} of an account, then use the approximate greedy algorithm to prune the original downstream network and build a minimum priority tree to speed up the iterative process. Through this process, the most suspicious subnetwork can be obtained as the core money laundering network.
%qs comment0202: 这里用过去时态好像有点歧义，可以考虑用can be obatined
%可疑度可以被替换为……参考frauder的叙述
The main idea of suspiciousness comes from the money laundering characteristics observed in RQ1 and RQ2: Money launderers will create a large and dense subgraph of transfers because money laundering accounts need to transfer a large number of funds in a short period of time to avoid being detected and frozen, resulting in a dense transfer subnetwork.

%qs comment0202:We define suspiciousness is... 好像有点奇怪，建议考虑改为 
%qs comment0202:f(s)公式中f_v(S)的v是什么？论文中没有介绍
The suspiciousness of subgraph $S=(N,E)$, where $N$ denotes nodes and $E$ denotes edges, is defined as: $$g(S)=\frac{f(S)}{|S|},$$
    which can be regarded as the result of taking the average value after summing the suspiciousness of each node and edge in the subgraph structure. Generally, \red{$f(S)$ is defined as follows:}
    \begin{equation*}
         \begin{aligned}
        f(S) & = f_N(S) + f_E(S) \\
            & = \sum_{i \in N} a_i + \sum_{i,j \in N,(i,j) \in E} c_{ij},
    \end{aligned}
    \end{equation*}
    where $a_i$ denotes node suspiciousness, $c_{ij}$ denotes edge suspiciousness.
    %qs comment0202:node suspiciousness $c_{ij}=0$ ?node sus好像是a_i? 这里这句话是不是多了$c_{ij}=0$ for edge suspiciousness. 感觉这一整句话有点怪
    %ai是节点怀疑度，cij是边怀疑度
    For simple calculation, we set edge suspiciousness $c_{ij}=0$ and $a_i = \alpha$ when node $i$ is labeled as a heist. In our experiment, we set $\alpha=49$ to achieve the best result according to the experiment.  Note that when it comes to application, the formula of $f(S)$ is general and can be replaced with other formulas according to needs.

%qs comment0202:这里的优先树相关的符号没有解释，如果木有时间加，建议删掉这段话，避免审稿人产生疑惑。另外句尾少符号了，我加了个.
%yym：把这一段挪到step一段里面了介绍优先树
% 洗钱核心网络识别算法：
%qs comment0202:The algorithm之前没重新开一个新段落，我直接换行开新段落啦

The algorithm\ref{alg:frauder} for identification of the money laundering core network is shown as follows:
% （1）子网络S的可疑度$g(S)=f(S)/|S|$, 可以理解成子网络结构中每个结点的可疑程度求和f(S)后取平均值的结果。
%qs comment0202:这里Step X之前没加 "：", 感觉加：会更合适一点？
\begin{enumerate}[Step 1:]
        
    \item Built a transaction network $G=(U \cup V,E)$, where $U$ is a set of transaction sender with size $m$, and $V$ is a set of recipients with size $n$, $E$ indicates transaction records.
    \item Build a priorities tree to restore the priorities $P$ of each node $i$, calculated as:
    $$P_i = f(X_t \textbackslash \{i\})-f(X_t).$$
    \item Traverse all nodes in $U$ and $V$, and calculate the suspiciousness after removing the current node $i^*$.
    \item Find out the subgraph with the largest suspiciousness.
    \item Update the priorities of nodes.
    \item Iterate accordingly until all nodes have been traversed.
    \item Get the subgraph $X^*$ with the largest suspiciousness.
    %qs comment0202:可以加一下迭代结束的条件嘛【yym：所有节点遍历完成】
    %遍历所有节点，计算去除节点后的怀疑度指标，找出怀疑度最大子图，更新权重，依此迭代，最终得到怀疑度最大子图
    %解释priorities
\end{enumerate}
%介绍算法符号，！！引用算法
% 【注】这里给出最终使用的可疑度指标公式：
% （2）基于最小优先树的近似贪心算法。
% 【注】模仿Fraudar算法写个伪代码，注意不要一模一样
\begin{algorithm}[t]
  \SetAlgoLined
  \KwData{Initially crawled Laundering Transaction newtwork $G=(U\cup V,E)$ by Algorithm 1, suspiciousness $g$ is defined before.} $X$ denotes the subgraph of $G$.
  \KwResult{A densest subgraph with maximum suspiciousness}
  initialization$(G,g)$\;
    construction of priority tree $T$ of $U\cup V$\;
    $X_0 \leftarrow U\cup V$\;
    \For{$t=1,...,m+n$}{
        $i^* \leftarrow argmax_{i\in X_i} g(X_i \textbackslash \{i\})$\;
        update the priorities in tree $T$ for all neighbors of $i^*$ \;
         $X_t \leftarrow X_{t-1} \textbackslash \{i^*\})$\;
    }
    $X^* \leftarrow argmax_{X_i\in \{X_0,...,X_{m+n}\} }$

  \caption{An approximate greedy algorithm based on a Minimal priority tree. }
  \label{alg:frauder}

\end{algorithm}
% 由于空间有限，我们这里仅展示了Upbit Hack洗钱案例的实验结果。根据Etherscan提供的标签，我们计算了洗钱核心网络中账户是否已被标记为洗钱的分类结果：Precision

Due to limited space, we only show the experimental process and results of the Upbit Hack money laundering case here. 
First, We built a bipartite graph based on the transaction data.
In the case of Upbit Hack, there are $131,654$ nodes in $U$, $16,138$ of which are labeled as heists, and $482,775$ nodes in $V$ with $16,536$ heists in them. The heists are marked according to the labels we get in the previous experiment. That is, we get a matrix of size $|U| \times |V| = 131,654 \times 482,775$. After implementing the algorithm, we get a core network of size $200 \times 1,332$. For the $200$ and $1,332$ nodes extracted from $U$ and $V$ respectively, we calculated the classification results of whether these account has been marked as money laundering: it turns out that the precision of our experiment reached $82.5\%$ and $100\%$ in $U$ and $V$ respectively. 
%基于上述可疑网络的发送方和接收方，我们可以得到一个可疑的交易子网络。原交易网络有2348180条交易数据，核心子网络的交易数据有xxxx条
Next, we collect the transaction data from our core network. The original transaction network has $2,348,180$ transactions while the core network we extract has only $45,811$ transactions.
% 同时，图*展示了通过如上算法得到的洗钱核心网络的可视化（用了Gephi的可视化工具）。
% 通过本文的这个方法，可以在原始的Upbit Hack洗钱网络的***个账户中，提取出包含了***个账户的洗钱核心网络，对于犯罪取证和调查工作来说，缩小了可疑的范围。

In all, through this method in this article, the money laundering core network of size $200 \times 1,332$ can be extracted from the original Upbit Hack money laundering network of \red{size} $16,138 \times 16,536$, which narrows the scope of suspicion for criminal evidence collection and investigation.
%qs comment0202:16,138和上面的数据我加了‘,’

\section{Ethical Considerations}
% 在本文中，我们第一次披露Web3洗钱数据集，调查并分析了黑客、攻击者、欺诈者等的洗钱手法。这可能会引起社会各界对犯罪模仿效应的担忧，但实际上，我们的研究动机和现有工作类似，如网络钓鱼诈骗检测[]、旁氏骗局、DApps攻击、伪造代币[]等。我们的数据集中公布的洗钱交易只是冰山一角。如本文所示，网络犯罪分子在Web3反洗钱的 "猫捉老鼠 "游戏中，正在逐年改进他们的方法和技术，这更加需要我们对加密货币洗钱进行调查和了解。本文工作根据我们在洗钱账户和网络中的有趣发现，可以促进更有效设计反洗钱算法，促进Web3生态系统的健康发展。

In this paper, we reveal the first crypto money laundering dataset in Web3, investigating and analyzing the money laundering techniques of hackers, exploiters, scammers, and others. The disclosure and investigation may cause the community to worry about contributing to the ``copycat crime'' effect, but actually, our research motivation is similar to the studies of Ponzi contracts~\cite{Ponzi2018Chen}, phishing scams~\cite{Li2021TTAGN}, DApp attacks~\cite{Su2021attacks}, counterfeit tokens~\cite{gao2020tracking}, etc. The money laundering transactions published in $\textsf{EthereumHeist}$ are only the tip of the iceberg. As shown in this paper, cybercriminals are improving their methods and techniques year by year in the ``cat-and-mouse'' game of Web3 anti-money laundering, reinforcing the need for investigation and understanding of crypto money laundering in the Web3 ecosystem. This work will facilitate more effective designs of anti-money laundering algorithms based on our interesting findings in the laundering accounts and networks, and further promote the healthy development of the Web3 ecosystem.
% , to facilitate more effective anti-money laundering algorithms and promote healthy development of the Web3 ecosystem
% 数据集发布出去以后，会不会有躲避洗钱检测
% 其实工业界很多年度洗钱报告，但目前还没有从学术的角度进行系统性的实证分析，从报告中可以看到，洗钱手段确实逐年愈加狡猾。目前暴露出来的其实也只是冰山一角。就像区块链欺诈检测一样，研究是为了更好的做反洗钱。确实存在一个对抗的趋势

% Add by Kaixin
% 至于我们的研究是否会涉及隐私问题，回答是否定的。首先，我们收集的数据是完全公开的，任何人都可以在区块链上获取。其次，我们数据集仅有链上的匿名交易数据，而没有和现实生活中的个人信息关联的其他数据。因此，基于以上两点，我们并不认为这项工作会侵犯到他人的隐私，也不会直接导致对个人的逮捕或起诉。
As for whether our research involves privacy issues, the answer is NO. First, the data we collect is completely public and can be accessed by anyone on blockchain. Second, our dataset only includes anonymous transaction data on blockchain, but not other data associated with real-life personal information. Therefore, based on these two points, we do not consider that this work will invade the privacy of others, or directly lead to the arrest or prosecution of individuals.

%-----------------------------------------------------------------------------------
%\vspace{-1ex}
\section{Discussion \& Conclusion}
% 合并原来的Section 8 and 9
In this paper, we conduct the \emph{first} systematic study to characterize the crypto money laundering in the Web3 ecosystem. We start from a very small number of security incident accounts, collect abundant money laundering transactions, and build a dataset named $\textsf{EthereumHeist}$.
%Based on the real-world data, we investigate the account features, network properties, and economic impact of crypto money laundering in the Web3 ecosystem. %, and have a more comprehensive understanding of the money laundering behaviors
Based on the dataset, we obtain a series of interesting findings of crypto money laundering in Web3 via answering three research questions from micro, meso to macro perspectives, reflecting the feasibility and necessity of Web3 AML. By answering RQ1 and RQ2, we summarize the characteristics at the account level and the network level, e.g. the lifespan and transaction amounts of accounts, and higher order patterns of sub-networks. These findings can help design effective red flag indicators and detective methods for Web3 AML.
% 纵观全文，我们从微观、中观到宏观的三个研究问题获得了Web3中加密货币洗钱的一些有趣发现，体现出Web3反洗钱的可行性和必要性。通过回答RQ1和RQ2，我们分别总结了洗钱账户层面和洗钱网络层面的特征，例如账户的寿命和交易金额，以及子网络的高阶模式。这些发现有助于为反洗钱系统设计有效的红旗指标。（方法层面，可行性）
Furthermore, by answering RQ3 in a data-driven manner, it can be observed that DApps on Web3 such as DEXes, lending services, etc. have been increasingly involved in money laundering activities in recent years. There is also evidence that dumping stolen money in the money laundering process affects price volatility. Therefore, money laundering is detrimental to the stability of the Web3 market, and it is necessary to develop decentralized security protocols based on economic incentives to achieve effective regulation of decentralized platforms in Web3. Coincidentally, the EU Commission has recently launched a public call for tender for a study on ``embedded supervision'' of DeFi~\cite{tender2022defi}. 

% 【Old version】The existence of multiple heists in this paper where accounts that have been dormant for several months resume their laundering activities indicates that stolen assets are still flowing in Web3 and are not fully regulated. Therefore, constructing red flag indicators based on money laundering behavior characteristics is a fundamental measure for AML in Web3. When the AML system monitors a large amount of illicit funds flowing into the Web3 environment, it should promptly analyze the transaction behavior of the relevant accounts, focusing on the transaction amount, lifespan, and higher-order characteristics among accounts, etc. 

% 另外，基于数据驱动的RQ3分析，可以看出基于DApp的洗钱行为由于其匿名性和监管不力而呈上升趋势，如DEXes、借贷服务、跨链平台等。也有证据表明，在洗钱过程中倾销赃款会影响价格波动。因此，洗钱不利于Web3市场的稳定，有必要制定基于经济激励的去中心化安全协议，以实现对Web3生态系统Dapp的有效监管。例如，欧盟委员会最近就以太坊上的DeFi的 "嵌入式监管 "研究进行了公开招标。

% 基于本文提出的包含大量节点和事件细节信息的Web3洗钱数据集和分析结果，未来研究者可以提出对于洗钱和诈骗的检测算法，正如那些基于椭圆数据集的检测方法一样。
% 
% 作为对Web3洗钱的初步研究，我们也存在一些局限性（在附录中讨论）为此，在我们的数据集和实证研究之后，未来的工作还有大量的机会。
As a preliminary exploration of Web3 money laundering and limited by space, this paper also has some limitations (discussed in the Appendix), and there are ample opportunities for future work following our dataset and analysis.
% (i) 由于本文提出的Web3洗钱数据集包含了大量节点，事件细节信息，以及分析结果，因此研究者有机会基于该数据集提出智能的洗钱账户追踪方法和洗钱团伙子网络检测算法，正如那些基于椭圆数据集的检测方法一样。
As the proposed dataset contains lots of accounts and transactions, and empirical results, researchers are able to propose intelligent tracing methods and money laundering subgraph detection based on this dataset, as those based on Elliptic dataset. %  mentioned in Section~\ref{subsec:aml}
Moreover, it is interesting to investigate the correlations and interactions between money laundering transactions in different cases, such as the hacks on Upbit and Kucoin, which are both responsible for Lazarus Group~\cite{Chainalysis2021}. This may provide further insights into the evolution of their money laundering strategies and lead to more accurate money laundering tracing. 
Last but not the least, researchers can also design strategies and methods based on cross-chain tracing in the Web3 ecosystem to extend the dataset of this paper.

\newpage

\bibliographystyle{ACM-Reference-Format}
\bibliography{main}


\begin{thebibliography}{38}

%%% the Web address.

\ifx \showCODEN    \undefined \def \showCODEN     #1{\unskip}     \fi
\ifx \showDOI      \undefined \def \showDOI       #1{#1}\fi
\ifx \showISBNx    \undefined \def \showISBNx     #1{\unskip}     \fi
\ifx \showISBNxiii \undefined \def \showISBNxiii  #1{\unskip}     \fi
\ifx \showISSN     \undefined \def \showISSN      #1{\unskip}     \fi
\ifx \showLCCN     \undefined \def \showLCCN      #1{\unskip}     \fi
\ifx \shownote     \undefined \def \shownote      #1{#1}          \fi
\ifx \showarticletitle \undefined \def \showarticletitle #1{#1}   \fi
\ifx \showURL      \undefined \def \showURL       {\relax}        \fi
% invisible to TeX
\providecommand\bibfield[2]{#2}
\providecommand\bibinfo[2]{#2}
\providecommand\natexlab[1]{#1}
\providecommand\showeprint[2][]{arXiv:#2}

\bibitem[\protect\citeauthoryear{Alarab, Prakoonwit, and Nacer}{Alarab
  et~al\mbox{.}}{2020}]%
        {alarab2020comparative}
\bibfield{author}{\bibinfo{person}{Ismail Alarab}, \bibinfo{person}{Simant
  Prakoonwit}, {and} \bibinfo{person}{Mohamed~Ikbal Nacer}.}
  \bibinfo{year}{2020}\natexlab{}.
\newblock \showarticletitle{Comparative analysis using supervised learning
  methods for anti-money laundering in bitcoin}. In
  \bibinfo{booktitle}{\emph{ICMLT}}. \bibinfo{pages}{11--17}.
\newblock


\bibitem[\protect\citeauthoryear{Benson, Gleich, and Leskovec}{Benson
  et~al\mbox{.}}{2016}]%
        {Benson2016Motif}
\bibfield{author}{\bibinfo{person}{Austin~R. Benson}, \bibinfo{person}{David~F.
  Gleich}, {and} \bibinfo{person}{Jure Leskovec}.}
  \bibinfo{year}{2016}\natexlab{}.
\newblock \showarticletitle{Higher-order organization of complex networks}.
\newblock \bibinfo{journal}{\emph{Science}} \bibinfo{volume}{353},
  \bibinfo{number}{6295} (\bibinfo{year}{2016}), \bibinfo{pages}{163--166}.
\newblock


\bibitem[\protect\citeauthoryear{Chen, Guo, Chen, Zheng, and Lu}{Chen
  et~al\mbox{.}}{2020}]%
        {Chen2020Phishing}
\bibfield{author}{\bibinfo{person}{Weili Chen}, \bibinfo{person}{Xiongfeng
  Guo}, \bibinfo{person}{Zhiguang Chen}, \bibinfo{person}{Zibin Zheng}, {and}
  \bibinfo{person}{Yutong Lu}.} \bibinfo{year}{2020}\natexlab{}.
\newblock \showarticletitle{Phishing scam detection on Ethereum: Towards
  financial security for blockchain ecosystem}. In
  \bibinfo{booktitle}{\emph{IJCAI}}. \bibinfo{pages}{4506--4512}.
\newblock


\bibitem[\protect\citeauthoryear{Chen, Wu, Zheng, Chen, and Zhou}{Chen
  et~al\mbox{.}}{2019}]%
        {chen2019market}
\bibfield{author}{\bibinfo{person}{Weili Chen}, \bibinfo{person}{Jun Wu},
  \bibinfo{person}{Zibin Zheng}, \bibinfo{person}{Chuan Chen}, {and}
  \bibinfo{person}{Yuren Zhou}.} \bibinfo{year}{2019}\natexlab{}.
\newblock \showarticletitle{Market manipulation of bitcoin: Evidence from
  mining the Mt. Gox transaction network}. In
  \bibinfo{booktitle}{\emph{INFOCOM}}. \bibinfo{pages}{964--972}.
\newblock


\bibitem[\protect\citeauthoryear{Chen, Zheng, Cui, Ngai, Zheng, and Zhou}{Chen
  et~al\mbox{.}}{2018}]%
        {Ponzi2018Chen}
\bibfield{author}{\bibinfo{person}{Weili Chen}, \bibinfo{person}{Zibin Zheng},
  \bibinfo{person}{Jiahui Cui}, \bibinfo{person}{Edith Ngai},
  \bibinfo{person}{Peilin Zheng}, {and} \bibinfo{person}{Yuren Zhou}.}
  \bibinfo{year}{2018}\natexlab{}.
\newblock \showarticletitle{Detecting ponzi schemes on Ethereum: Towards
  healthier blockchain technology}. In \bibinfo{booktitle}{\emph{WWW}}.
  \bibinfo{pages}{1409–1418}.
\newblock


\bibitem[\protect\citeauthoryear{Commission}{Commission}{2022}]%
        {tender2022defi}
\bibfield{author}{\bibinfo{person}{European~Union Commission}.}
  \bibinfo{year}{2022}\natexlab{}.
\newblock \bibinfo{booktitle}{\emph{Study on Embedded Supervision of
  Decentralised Finance 2022/S 192-542418 Contract Notice}}.
\newblock
\urldef\tempurl%
\url{https://ted.europa.eu/udl?uri=TED:NOTICE:542418-2022:HTML:EN:HTML&tabId=1&tabLang=en}
\showURL{%
\tempurl}


\bibitem[\protect\citeauthoryear{Company}{Company}{2022}]%
        {McKinsey2022Web3}
\bibfield{author}{\bibinfo{person}{McKinsey~\& Company}.}
  \bibinfo{year}{2022}\natexlab{}.
\newblock \bibinfo{booktitle}{\emph{Web3 beyond the hype}}.
\newblock
\urldef\tempurl%
\url{https://www.mckinsey.com/industries/financial-services/our-insights/web3-beyond-the-hype}
\showURL{%
\tempurl}


\bibitem[\protect\citeauthoryear{Dumitrescu, Băltoiu, and Budulan}{Dumitrescu
  et~al\mbox{.}}{2022}]%
        {Dumitrescu2022Bank}
\bibfield{author}{\bibinfo{person}{Bogdan Dumitrescu}, \bibinfo{person}{Andra
  Băltoiu}, {and} \bibinfo{person}{Ştefania Budulan}.}
  \bibinfo{year}{2022}\natexlab{}.
\newblock \showarticletitle{Anomaly detection in graphs of bank transactions
  for anti money laundering applications}.
\newblock \bibinfo{journal}{\emph{IEEE Access}}  \bibinfo{volume}{10}
  (\bibinfo{year}{2022}), \bibinfo{pages}{47699--47714}.
\newblock


\bibitem[\protect\citeauthoryear{FATF}{FATF}{2019}]%
        {fatf2019VASP_old}
\bibfield{author}{\bibinfo{person}{FATF}.} \bibinfo{year}{2019}\natexlab{}.
\newblock \bibinfo{booktitle}{\emph{Guidance for a risk-based approach to
  virtual assets and virtual asset service providers}}.
\newblock
\urldef\tempurl%
\url{www.fatf-gafi.org/publications/fatfrecommendations/documents/Guidance-RBA-virtual-assets.html}
\showURL{%
\tempurl}


\bibitem[\protect\citeauthoryear{FATF}{FATF}{2021}]%
        {fatf2021VASP}
\bibfield{author}{\bibinfo{person}{FATF}.} \bibinfo{year}{2021}\natexlab{}.
\newblock \bibinfo{booktitle}{\emph{Updated guidance for a risk-based approach
  to virtual assets and virtual asset service providers}}.
\newblock
\urldef\tempurl%
\url{www.fatf-gafi.org/publications/fatfrecommendations/documents/Updated-Guidance-RBA-VA-VASP.html}
\showURL{%
\tempurl}


\bibitem[\protect\citeauthoryear{FATF}{FATF}{2022}]%
        {fatf2022}
\bibfield{author}{\bibinfo{person}{FATF}.} \bibinfo{year}{2022}\natexlab{}.
\newblock \bibinfo{booktitle}{\emph{Money laundering and terrorist financing
  red flag indicators associated with virtual assets}}.
\newblock
\urldef\tempurl%
\url{http://www.fatf-gafi.org/publications/fatfrecommendations/documents/Virtual-Assets-Red-Flag-Indicators.html}
\showURL{%
\tempurl}


\bibitem[\protect\citeauthoryear{Gao, Wang, Xia, Wu, Zhou, Luo, and Tyson}{Gao
  et~al\mbox{.}}{2020}]%
        {gao2020tracking}
\bibfield{author}{\bibinfo{person}{Bingyu Gao}, \bibinfo{person}{Haoyu Wang},
  \bibinfo{person}{Pengcheng Xia}, \bibinfo{person}{Siwei Wu},
  \bibinfo{person}{Yajin Zhou}, \bibinfo{person}{Xiapu Luo}, {and}
  \bibinfo{person}{Gareth Tyson}.} \bibinfo{year}{2020}\natexlab{}.
\newblock \showarticletitle{Tracking counterfeit cryptocurrency end-to-end}.
\newblock \bibinfo{journal}{\emph{SIGMERICS}} \bibinfo{volume}{4},
  \bibinfo{number}{3}, \bibinfo{pages}{1--28}.
\newblock


\bibitem[\protect\citeauthoryear{Gao and Ye}{Gao and Ye}{2007}]%
        {Gao2007data}
\bibfield{author}{\bibinfo{person}{Zengan Gao} {and} \bibinfo{person}{Mao Ye}.}
  \bibinfo{year}{2007}\natexlab{}.
\newblock \showarticletitle{A framework for data mining-based anti-money
  laundering research}.
\newblock \bibinfo{journal}{\emph{Journal of Money Laundering Control}}
  \bibinfo{volume}{10}, \bibinfo{number}{2} (\bibinfo{year}{2007}),
  \bibinfo{pages}{170--179}.
\newblock


\bibitem[\protect\citeauthoryear{Kolachala, Simsek, Ababneh, and
  Vishwanathan}{Kolachala et~al\mbox{.}}{2021}]%
        {Kolachala2021SoK}
\bibfield{author}{\bibinfo{person}{Kartick Kolachala}, \bibinfo{person}{Ecem
  Simsek}, \bibinfo{person}{Mohammed Ababneh}, {and} \bibinfo{person}{Roopa
  Vishwanathan}.} \bibinfo{year}{2021}\natexlab{}.
\newblock \showarticletitle{{SoK: Money laundering in cryptocurrencies}}. In
  \bibinfo{booktitle}{\emph{ARES}}.
\newblock


\bibitem[\protect\citeauthoryear{Lee, Khan, {Sen Gupta}, Ong, and Liu}{Lee
  et~al\mbox{.}}{2020}]%
        {Lee2020Measurements}
\bibfield{author}{\bibinfo{person}{Xi~Tong Lee}, \bibinfo{person}{Arijit Khan},
  \bibinfo{person}{Sourav {Sen Gupta}}, \bibinfo{person}{Yu~Hann Ong}, {and}
  \bibinfo{person}{Xuan Liu}.} \bibinfo{year}{2020}\natexlab{}.
\newblock \showarticletitle{{Measurements, analyses, and insights on the entire
  Ethereum blockchain network}}. In \bibinfo{booktitle}{\emph{WWW}}.
  \bibinfo{pages}{155--166}.
\newblock


\bibitem[\protect\citeauthoryear{Li, Gou, Liu, Hou, Li, and Xiong}{Li
  et~al\mbox{.}}{2021}]%
        {Li2021TTAGN}
\bibfield{author}{\bibinfo{person}{Sijia Li}, \bibinfo{person}{Gaopeng Gou},
  \bibinfo{person}{Chang Liu}, \bibinfo{person}{Chengshang Hou},
  \bibinfo{person}{Zhenzhen Li}, {and} \bibinfo{person}{Gang Xiong}.}
  \bibinfo{year}{2021}\natexlab{}.
\newblock \showarticletitle{{TTAGN : Temporal transaction aggregation graph
  network for Ethereum phishing scams detection}}. In
  \bibinfo{booktitle}{\emph{WWW}}. \bibinfo{pages}{661--669}.
\newblock


\bibitem[\protect\citeauthoryear{Lorenz, Silva, Apar{\'\i}cio, Ascens{\~a}o,
  and Bizarro}{Lorenz et~al\mbox{.}}{2020}]%
        {lorenz2020machine}
\bibfield{author}{\bibinfo{person}{Joana Lorenz},
  \bibinfo{person}{Maria~In{\^e}s Silva}, \bibinfo{person}{David
  Apar{\'\i}cio}, \bibinfo{person}{Jo{\~a}o~Tiago Ascens{\~a}o}, {and}
  \bibinfo{person}{Pedro Bizarro}.} \bibinfo{year}{2020}\natexlab{}.
\newblock \showarticletitle{Machine learning methods to detect money laundering
  in the bitcoin blockchain in the presence of label scarcity}. In
  \bibinfo{booktitle}{\emph{ICAIF}}.
\newblock


\bibitem[\protect\citeauthoryear{M{\"o}ser, B{\"o}hme, and Breuker}{M{\"o}ser
  et~al\mbox{.}}{2014}]%
        {risk2014Malte}
\bibfield{author}{\bibinfo{person}{Malte M{\"o}ser}, \bibinfo{person}{Rainer
  B{\"o}hme}, {and} \bibinfo{person}{Dominic Breuker}.}
  \bibinfo{year}{2014}\natexlab{}.
\newblock \showarticletitle{Towards risk scoring of Bitcoin transactions}. In
  \bibinfo{booktitle}{\emph{FC}}. \bibinfo{pages}{16--32}.
\newblock


\bibitem[\protect\citeauthoryear{Rocha-Salazar, Segovia-Vargas, and
  Camacho-Mi{\~{n}}ano}{Rocha-Salazar et~al\mbox{.}}{2021}]%
        {Rocha-Salazar2021}
\bibfield{author}{\bibinfo{person}{Jos{\'{e}} de~Jes{\'{u}}s Rocha-Salazar},
  \bibinfo{person}{Mar{\'{i}}a~Jes{\'{u}}s Segovia-Vargas}, {and}
  \bibinfo{person}{Mar{\'{i}}a del~Mar Camacho-Mi{\~{n}}ano}.}
  \bibinfo{year}{2021}\natexlab{}.
\newblock \showarticletitle{Money laundering and terrorism financing detection
  using neural networks and an abnormality indicator}.
\newblock \bibinfo{journal}{\emph{Expert Systems with Applications}}
  \bibinfo{volume}{169} (\bibinfo{year}{2021}), \bibinfo{pages}{114470}.
\newblock


\bibitem[\protect\citeauthoryear{{Siwei Wu}, {Dabao Wang}, {Jianting He},
  {Yajin Zhou}, {Lei Wu}, {Xingliang Yuan}, {Qinming He}, and {Kui Ren}}{{Siwei
  Wu} et~al\mbox{.}}{2021}]%
        {Wu2021DeFiRanger}
\bibfield{author}{\bibinfo{person}{{Siwei Wu}}, \bibinfo{person}{{Dabao Wang}},
  \bibinfo{person}{{Jianting He}}, \bibinfo{person}{{Yajin Zhou}},
  \bibinfo{person}{{Lei Wu}}, \bibinfo{person}{{Xingliang Yuan}},
  \bibinfo{person}{{Qinming He}}, {and} \bibinfo{person}{{Kui Ren}}.}
  \bibinfo{year}{2021}\natexlab{}.
\newblock \showarticletitle{{DeFiRanger}: {D}etecting price manipulation
  attacks on DeFi applications}.
\newblock \bibinfo{journal}{\emph{arXiv preprint arXiv:2104.15068}}
  (\bibinfo{year}{2021}).
\newblock


\bibitem[\protect\citeauthoryear{Su, Shen, Du, Liao, Wang, Xing, and Liu}{Su
  et~al\mbox{.}}{2021}]%
        {Su2021attacks}
\bibfield{author}{\bibinfo{person}{Liya Su}, \bibinfo{person}{Xinyue Shen},
  \bibinfo{person}{Xiangyu Du}, \bibinfo{person}{Xiaojing Liao},
  \bibinfo{person}{XiaoFeng Wang}, \bibinfo{person}{Luyi Xing}, {and}
  \bibinfo{person}{Baoxu Liu}.} \bibinfo{year}{2021}\natexlab{}.
\newblock \showarticletitle{{Evil under the sun: Understanding and discovering
  attacks on Ethereum decentralized applications}}. In
  \bibinfo{booktitle}{\emph{USENIX Security}}. \bibinfo{pages}{1307--1324}.
\newblock


\bibitem[\protect\citeauthoryear{Team}{Team}{2021}]%
        {Chainalysis2021}
\bibfield{author}{\bibinfo{person}{Chainalysis Team}.}
  \bibinfo{year}{2021}\natexlab{}.
\newblock \bibinfo{booktitle}{\emph{Report: Key players of the cryptocurrency
  ecosystem}}.
\newblock
\urldef\tempurl%
\url{https://go.chainalysis.com/rs/503-FAP-074/images/Key-players-in-crypto-report.pdf}
\showURL{%
\tempurl}


\bibitem[\protect\citeauthoryear{Team}{Team}{2022a}]%
        {Chainalysis2022}
\bibfield{author}{\bibinfo{person}{Chainalysis Team}.}
  \bibinfo{year}{2022}\natexlab{a}.
\newblock \bibinfo{booktitle}{\emph{The Chainalysis 2022 Crypto Crime Report}}.
\newblock
\urldef\tempurl%
\url{https://go.chainalysis.com/2022-crypto-crime-report.html}
\showURL{%
\tempurl}


\bibitem[\protect\citeauthoryear{Team}{Team}{2022b}]%
        {Chainalysis2022Web3}
\bibfield{author}{\bibinfo{person}{Chainalysis Team}.}
  \bibinfo{year}{2022}\natexlab{b}.
\newblock \bibinfo{booktitle}{\emph{The chainalysis state of web3 report}}.
\newblock
\urldef\tempurl%
\url{https://go.chainalysis.com/2022-web3-report.html}
\showURL{%
\tempurl}


\bibitem[\protect\citeauthoryear{Team}{Team}{2022c}]%
        {CipherTrace2022}
\bibfield{author}{\bibinfo{person}{CipherTrace Team}.}
  \bibinfo{year}{2022}\natexlab{c}.
\newblock \bibinfo{booktitle}{\emph{Cryptocurrency crime and anti-money
  laundering report}}.
\newblock
\urldef\tempurl%
\url{https://ciphertrace.com/crime-and-anti-money-laundering-report-october-2022/}
\showURL{%
\tempurl}


\bibitem[\protect\citeauthoryear{Team}{Team}{2022d}]%
        {certik2022Web3}
\bibfield{author}{\bibinfo{person}{CertiK Team}.}
  \bibinfo{year}{2022}\natexlab{d}.
\newblock \bibinfo{booktitle}{\emph{The web3 security quarterly report (Q2
  2022)}}.
\newblock
\urldef\tempurl%
\url{https://certik-2.hubspotpagebuilder.com/hack3d-q1-2022-0}
\showURL{%
\tempurl}


\bibitem[\protect\citeauthoryear{Team}{Team}{2022e}]%
        {slowmist2022}
\bibfield{author}{\bibinfo{person}{SlowMist Team}.}
  \bibinfo{year}{2022}\natexlab{e}.
\newblock \bibinfo{booktitle}{\emph{First half of the 2022 blockchain security
  and anti-money laundering analysis report}}.
\newblock
\urldef\tempurl%
\url{https://www.slowmist.com/report/first-half-of-the-2022-report.pdf}
\showURL{%
\tempurl}


\bibitem[\protect\citeauthoryear{Turner, McCombie, and Uhlmann}{Turner
  et~al\mbox{.}}{2020}]%
        {turner2020discerning}
\bibfield{author}{\bibinfo{person}{Adam~B Turner}, \bibinfo{person}{Stephen
  McCombie}, {and} \bibinfo{person}{Allon~J Uhlmann}.}
  \bibinfo{year}{2020}\natexlab{}.
\newblock \showarticletitle{Discerning payment patterns in Bitcoin from
  ransomware attacks}.
\newblock \bibinfo{journal}{\emph{Journal of Money Laundering Control}}
  \bibinfo{volume}{23}, \bibinfo{number}{3} (\bibinfo{year}{2020}),
  \bibinfo{pages}{545--589}.
\newblock


\bibitem[\protect\citeauthoryear{Victor and Weintraud}{Victor and
  Weintraud}{2021}]%
        {victor2021detecting}
\bibfield{author}{\bibinfo{person}{Friedhelm Victor} {and}
  \bibinfo{person}{Andrea~Marie Weintraud}.} \bibinfo{year}{2021}\natexlab{}.
\newblock \showarticletitle{Detecting and quantifying wash trading on
  decentralized cryptocurrency exchanges}. In \bibinfo{booktitle}{\emph{WWW}}.
  \bibinfo{pages}{23--32}.
\newblock


\bibitem[\protect\citeauthoryear{Weber, Domeniconi, Chen, Weidele, Bellei,
  Robinson, and Leiserson}{Weber et~al\mbox{.}}{2019}]%
        {elliptic2019Weber}
\bibfield{author}{\bibinfo{person}{Mark Weber}, \bibinfo{person}{Giacomo
  Domeniconi}, \bibinfo{person}{Jie Chen}, \bibinfo{person}{Daniel Karl~I.
  Weidele}, \bibinfo{person}{Claudio Bellei}, \bibinfo{person}{Tom Robinson},
  {and} \bibinfo{person}{Charles~E. Leiserson}.}
  \bibinfo{year}{2019}\natexlab{}.
\newblock \showarticletitle{Anti-money laundering in Bitcoin: Experimenting
  with graph convolutional networks for financial forensics}.
\newblock \bibinfo{journal}{\emph{arXiv preprint arXiv:1908.02591}}
  (\bibinfo{year}{2019}).
\newblock


\bibitem[\protect\citeauthoryear{Werner, Perez, Gudgeon, Klages-Mundt, Harz,
  and Knottenbelt}{Werner et~al\mbox{.}}{2021}]%
        {werner2021sok}
\bibfield{author}{\bibinfo{person}{Sam~M Werner}, \bibinfo{person}{Daniel
  Perez}, \bibinfo{person}{Lewis Gudgeon}, \bibinfo{person}{Ariah
  Klages-Mundt}, \bibinfo{person}{Dominik Harz}, {and}
  \bibinfo{person}{William~J Knottenbelt}.} \bibinfo{year}{2021}\natexlab{}.
\newblock \showarticletitle{Sok: Decentralized finance (defi)}.
\newblock \bibinfo{journal}{\emph{arXiv preprint arXiv:2101.08778}}
  (\bibinfo{year}{2021}).
\newblock


\bibitem[\protect\citeauthoryear{Wronka}{Wronka}{2021}]%
        {Wronka2021Cyber}
\bibfield{author}{\bibinfo{person}{Christoph Wronka}.}
  \bibinfo{year}{2021}\natexlab{}.
\newblock \showarticletitle{{“Cyber-laundering”: The change of money
  laundering in the digital age}}.
\newblock \bibinfo{journal}{\emph{Journal of Money Laundering Control}}
  \bibinfo{volume}{25}, \bibinfo{number}{2} (\bibinfo{year}{2021}),
  \bibinfo{pages}{330--344}.
\newblock


\bibitem[\protect\citeauthoryear{Wu, Liu, Zhao, and Zheng}{Wu
  et~al\mbox{.}}{2021b}]%
        {Wu2021JNCA}
\bibfield{author}{\bibinfo{person}{Jiajing Wu}, \bibinfo{person}{Jieli Liu},
  \bibinfo{person}{Yijing Zhao}, {and} \bibinfo{person}{Zibin Zheng}.}
  \bibinfo{year}{2021}\natexlab{b}.
\newblock \showarticletitle{{Analysis of cryptocurrency transactions from a
  network perspective: An overview}}.
\newblock \bibinfo{journal}{\emph{Journal of Network and Computer
  Applications}}  \bibinfo{volume}{190} (\bibinfo{year}{2021}),
  \bibinfo{pages}{103139}.
\newblock


\bibitem[\protect\citeauthoryear{Wu, Yuan, Lin, You, Chen, Chen, and Zheng}{Wu
  et~al\mbox{.}}{2022b}]%
        {trans2vec2020Wu}
\bibfield{author}{\bibinfo{person}{Jiajing Wu}, \bibinfo{person}{Qi Yuan},
  \bibinfo{person}{Dan Lin}, \bibinfo{person}{Wei You}, \bibinfo{person}{Weili
  Chen}, \bibinfo{person}{Chuan Chen}, {and} \bibinfo{person}{Zibin Zheng}.}
  \bibinfo{year}{2022}\natexlab{b}.
\newblock \showarticletitle{Who are the phishers? Phishing scam detection on
  Ethereum via network embedding}.
\newblock \bibinfo{journal}{\emph{IEEE Transactions on Systems, Man, and
  Cybernetics: Systems}} \bibinfo{volume}{52}, \bibinfo{number}{2}
  (\bibinfo{year}{2022}), \bibinfo{pages}{1156--1166}.
\newblock


\bibitem[\protect\citeauthoryear{Wu, Hu, Zhou, Wang, Luo, Wang, Zhang, and
  Ren}{Wu et~al\mbox{.}}{2021a}]%
        {Wu2021Mixing}
\bibfield{author}{\bibinfo{person}{Lei Wu}, \bibinfo{person}{Yufeng Hu},
  \bibinfo{person}{Yajin Zhou}, \bibinfo{person}{Haoyu Wang},
  \bibinfo{person}{Xiapu Luo}, \bibinfo{person}{Zhi Wang}, \bibinfo{person}{Fan
  Zhang}, {and} \bibinfo{person}{Kui Ren}.} \bibinfo{year}{2021}\natexlab{a}.
\newblock \showarticletitle{Towards understanding and semystifying bitcoin
  mixing services}. In \bibinfo{booktitle}{\emph{WWW}}.
  \bibinfo{pages}{33--44}.
\newblock


\bibitem[\protect\citeauthoryear{Wu, Liu, Wu, and Zheng}{Wu
  et~al\mbox{.}}{2022a}]%
        {wu2022transaction}
\bibfield{author}{\bibinfo{person}{Zhiying Wu}, \bibinfo{person}{Jieli Liu},
  \bibinfo{person}{Jiajing Wu}, {and} \bibinfo{person}{Zibin Zheng}.}
  \bibinfo{year}{2022}\natexlab{a}.
\newblock \showarticletitle{Transaction tracking on blockchain trading systems
  using personalized PageRank}.
\newblock \bibinfo{journal}{\emph{arXiv preprint arXiv:2201.05757}}
  (\bibinfo{year}{2022}).
\newblock


\bibitem[\protect\citeauthoryear{Xia, Wang, Gao, Su, Yu, Luo, Zhang, Xiao, and
  Xu}{Xia et~al\mbox{.}}{2021}]%
        {xia2021trade}
\bibfield{author}{\bibinfo{person}{Pengcheng Xia}, \bibinfo{person}{Haoyu
  Wang}, \bibinfo{person}{Bingyu Gao}, \bibinfo{person}{Weihang Su},
  \bibinfo{person}{Zhou Yu}, \bibinfo{person}{Xiapu Luo}, \bibinfo{person}{Chao
  Zhang}, \bibinfo{person}{Xusheng Xiao}, {and} \bibinfo{person}{Guoai Xu}.}
  \bibinfo{year}{2021}\natexlab{}.
\newblock \showarticletitle{Trade or trick? Detecting and characterizing scam
  tokens on Uniswap decentralized exchange}. In
  \bibinfo{booktitle}{\emph{SIGMERICS}}, Vol.~\bibinfo{volume}{5}.
\newblock


\bibitem[\protect\citeauthoryear{Zheng, Zheng, Wu, and Dai}{Zheng
  et~al\mbox{.}}{2020}]%
        {Zheng2020XBlockETH}
\bibfield{author}{\bibinfo{person}{Peilin Zheng}, \bibinfo{person}{Zibin
  Zheng}, \bibinfo{person}{Jiajing Wu}, {and} \bibinfo{person}{Hong-Ning Dai}.}
  \bibinfo{year}{2020}\natexlab{}.
\newblock \showarticletitle{XBlock-ETH: {E}xtracting and exploring blockchain
  data From Ethereum}.
\newblock  \bibinfo{volume}{1}, \bibinfo{number}{April} (\bibinfo{year}{2020}),
  \bibinfo{pages}{95--106}.
\newblock


\end{thebibliography}
%\bibliography{ref-0904}

\setcitestyle{numbers,sort&compress}

\newpage

%newpage

\newpage

\end{document}